\documentclass[usenatbib]{mn2e}
\usepackage{times}
\usepackage{graphicx}
\usepackage{epsfig}
\usepackage{amsfonts}
\usepackage{amssymb}
\usepackage{rotating} 
\def\apj{{\em ApJ}}
\def\apjs{{\em ApJS}}
\def\apjl{{\em ApJ}}
\def\aap{{\em A\&A}}
\def\aj{{\em AJ}}

\def\mnras{{\em MNRAS}}

\def\beq{\begin{equation}}
\def\eeq{\end{equation}}
\title[Topology and Sizes of H~II Regions during Cosmic Reionization] 
{Topology and Sizes of H~II Regions during Cosmic Reionization}
\author[Friedrich et al.]{Martina M. Friedrich$^{1}$
\thanks{e--mail: martina@astro.su.se}, 
Garrelt Mellema$^1$,
Marcelo A. Alvarez$^{2}$,
Paul R. Shapiro$^{3}$ and 
\newauthor 
Ilian T. Iliev$^{4}$\\
$^1$Department of Astronomy \& Oskar Klein Centre, AlbaNova, Stockholm University, SE-106 91
Stockholm, Sweden \\
$^2$Canadian Institute for Theoretical Astrophysics, University of
Toronto, 60 St. George Street, Toronto, ON M5S 3H8, Canada\\
$^3$Department of Astronomy and the Texas Cosmology Center, The University of 
Texas at Austin, TX 78712, USA\\
$^4$Astronomy Centre, Department of Physics \& Astronomy, Pevensey II Building, 
University of Sussex, Falmer, Brighton BN1 9QH\\
}
\date{\today}
\pubyear{2010} \volume{001} \pagerange{1}
\begin{document}
\pagerange{\pageref{firstpage}--\pageref{lastpage}}
\maketitle
\label{firstpage}
%-------------------------------------------------------------------------------
%-------------------------------------------------------------------------------
% ABSTRACT----------------------------------------------------------------------
\begin{abstract}
  We use the results of large-scale simulations of reionization to
  explore methods for characterizing the topology and sizes of HII
  regions during reionization. 
  We use four independent methods for characterizing the sizes of
  ionized regions. Three of them give us a full size distribution: the
  friends-of-friends (FOF) method, the spherical average method (SPA)
  and the power spectrum (PS) of
  the ionized fraction. These latter three methods are complementary:
  While the
  FOF method captures the size distribution of the small scale H~II
  regions, which contribute only a small amount to the total ionization
  fraction, the spherical average method provides a smoothed measure
  for the average size of the H~II regions constituting the main
  contribution to the ionized fraction, and the power spectrum does
  the same while retaining more details on the size distribution.
  Our fourth method for characterizing the sizes of the H II
  regions is the average size which results if we divide
  the total volume of the H II regions by their total surface area,
  (i.e. 3V/A), computed in terms of the ratio of the corresponding
  Minkowski functionals of the ionized fraction field.
  To characterize the
  topology of the ionized regions, we calculate the evolution of the
  Euler Characteristic. We find that the evolution of the topology
  during the first half of reionization is consistent with inside-out
  reionization of a Gaussian density field. We use these techniques to
  investigate the dependence of size and topology on some basic source
  properties, such as the halo mass-to-light ratio, susceptibility of
  haloes to negative feedback from reionization, and the minimum halo
  mass for sources to form. We find that suppression of ionizing
  sources within ionized regions slows the growth of H~II regions, and also
  changes their size distribution. Additionally, the topology of
  simulations including suppression is more complex, as indicated by
  the evolution of the Euler characteristic of the ionized regions.
  We find density and ionized fraction to be correlated on large
  scales, in agreement with the inside-out picture of reionization.
\end{abstract}
%-------------------------------------------------------------------------------
%-------------------------------------------------------------------------------
%-------------------------------------------------------------------------------
% KEYWORDS----------------------------------------------------------------------
%-------------------------------------------------------------------------------
%-------------------------------------------------------------------------------
%-------------------------------------------------------------------------------
% INTRODUCTION------------------------------------------------------------------
\section{Introduction} 
The Cosmic Microwave Background (CMB) discovered in 1965 was
evidence that the hot big bang universe cooled and recombined
\citep{1965ApJ...142..419P}.
That same year, however, the intergalactic medium (IGM) at z = 2 was found to be 
largely devoid of neutral hydrogen atoms, when
astronomers failed to detect their Lyman alpha resonant scattering
in the spectra of the first quasars discovered with high enough redshift
to make the transition visible from the ground
\citep{1965ApJ...142.1633G, 1966ApJ...145..668O}.  This was soon interpreted to mean
that, unless the IGM were many orders of magnitude less dense than
the average density of a critical universe, most of the
hydrogen atoms there must have been reionized sometime between $z = 1000$
and $z = 2$.

Although we have since then learned much more about both the CMB and
the HI absorption towards high redshift QSOs, currently it is still
those two observables which constrain the epoch of reionization
(EoR). The results from the WMAP measurements of the CMB have
constrained the optical depth due to electron scattering,
$\tau_\mathrm{es}$, to $0.088 \pm 0.015$, implying that an
instantaneous reionization would have happened at $z=10.4\pm 1.2$
\citep{2010arXiv1001.4538K}. The QSO spectra obtained within the Sloan
Digital Sky Survey (SDSS) indicate a low, but rapidly rising neutral
fraction around redshift 6 \citep{2006AJ....132..117F,
  2007AJ....134.2435W}. The combination of these two measurements
suggests that the epoch of reionization extended over several redshift
units.

However, a series of measurements are being performed or prepared that
are expected to give completely new constraints on this early epoch of
galaxy formation.  Radio telescopes capable of measuring at low
frequencies (GMRT\footnote{Giant Metrewave Telescope,
  http://gmrt.ncra.tifr.res.in}, 21CMA\footnote{21 Centimeter Array,
  http://21cma.bao.ac.cn}, LOFAR\footnote{Low Frequency Array,
  http://www.lofar.org}, MWA\footnote{Murchison Widefield Array,
  http://www.mwatelescope.org}, PAPER\footnote{Precision Array to
  Probe the EoR, http://astro.berkeley.edu/\~{}dbacker/eor}) should be able
to detect the signature of redshifted 21cm radiation from neutral
hydrogen during the EoR. These measurements are challenging due to the
presence of the strong foreground emission of, mostly, our own Milky
Way as well as ionospheric distortions. If successful, these
experiments should produce a detection before the 50th anniversary of
the discovery of the CMB and the Gunn-Peterson effect.

In preparation for these 21cm observations, many groups have been
numerically simulating the reionization process on large scales
\citep[e.g.][ to name a small selection]{iliev06,
  2007MNRAS.377.1043M, 2008ApJ...681..756S}; see
\citet{2009arXiv0906.4348T} for a review on simulations of
reionization. Semi-numerical models have also been developed, first by
\citet{2007ApJ...654...12Z}, later improved by others
\citep{2007ApJ...669..663M, 2008ApJ...689....1S, 2009ApJ...703L.167A,
 2009MNRAS.394..960C, 2010arXiv1003.3878M, 2010arXiv1003.6132A}. 
What both of these types of calculations give
us is the evolution of the ionized fraction of intergalactic hydrogen in the
Universe, $x(\mathbf{r},t)$.

The simulation results show a great amount of complexity in $x(\mathbf{r},t)$. As
the sources of reionization are likely to be clustered in space,
individual H~II regions typically contain many sources
\citep[e.g.,][]{2004ApJ...613....1F,2005ApJ...624..491I,
  2006MNRAS.369.1625I} and obtain complex shapes in 3D space. Rare
sources are more biased than more abundant ones, and it is expected
that the level of bias will largely determine the characteristic scale
of the reionization process \citep{2006MNRAS.369.1625I,
  2006MNRAS.365..115F}. Accurate theoretical predictions for the
morphology and size of H~II regions depend upon an understanding of
the the abundance and clustering of the ionizing sources themselves,
in addition to the underlying inhomogeneous density
field. Quantitative analysis of the distribution of ionized material
during the EoR is thus not a trivial matter and likely several
different approaches have to be combined. The main aim of this paper is to
describe and evaluate different methods for analyzing the properties
of the ionization fraction field $x(\mathbf{r},t)$ and its evolution.

The planned observations of the 21cm line of neutral hydrogen are
expected to constrain the ionization fraction field 
$x(\mathbf{r},t)$ statistically as the power
spectrum of neutral hydrogen fluctuations is the most directly
observed quantity via 21--cm radio observations
\citep[e.g.][]{2004ApJ...608..622Z, mellema06,
  2010arXiv1003.0965H}. Future observations, for example with the SKA
(Square Kilometer Array) may have enough sensitivity to actually image
the redshifted 21cm signal as a function of frequency and thus reveal the
spatial structure of the ionization fraction field $x(\mathbf{r},t)$.

Ultimately, we are less interested in the function $x(\mathbf{r},t)$ itself but more
in ``why it is like it is'', i.e. in the properties of the sources and 
sinks of reionization. For this, one has to find out how the
statistical properties of ionization fraction field depend on different source and sink
properties. Simulations of reionization can thus be said not to aim at reproducing
the actual $x(\mathbf{r},t)$, but rather at showing the same statistical behavior as the real
epoch of reionization: on the one hand, because only statistical
quantities can be measured (as mentioned above), on the other hand
because the input of simulations is only statistically comparable to
the real conditions. 

This paper has two parts. In the first part we investigate the usefulness,
in terms of characterizing size distributions of ionized regions, of
different kinds of statistics (for example the power spectrum of
$x(\mathbf{r},t)$) of a simulated ionization fraction field. In the second part we employ these
statistics to investigate the effect of different source properties.
This is useful to draw conclusions on the sources, once statistical
properties of the real $x(\mathbf{r},t)$ of reionization can be measured.

We focus on the early and intermediate stages of reionization, when
the morphology of H~II regions is most well defined and the photon
mean-free-path is determined by the patchiness of the reionization
process itself. At the latest stages of the reionization process,
after overlap, fluctuations in the UV background are expected to be
sensitive to the small fraction of gas which is left neutral in the
form, for example, of Lyman-limit systems
\citep{2000ApJ...530....1M,2006ApJ...648....1G, 2010arXiv1003.6132A, 
2009arXiv0912.0292P}. We limit ourselves to
analyzing the ionization fraction fields $x(\mathbf{r},t)$ from the simulations, not
on producing the observable quantities. This is the necessary first
step before proceeding to evaluate whether different scenarios can be
observationally distinguished. The observables will be discussed in 
a follow-up paper (Iliev et al., in prep.). 

In terms of sections, the paper is organized as follows: \S2
introduces the simulations included in this study. \S3 introduces the
analysis methods used to investigate these simulations. In \S4 we test
the effect of numerical parameters on the statistics of $x(\mathbf{r},t)$. In \S5,
we test the effect of source properties on the statistics of $x(\mathbf{r},t)$. We
end with our conclusions in \S6.
%-------------------------------------------------------------------------------
%-------------------------------------------------------------------------------
%-------------------------------------------------------------------------------
% SIMULATIONS-------------------------------------------------------------------
\section{Simulations} 

\label{sims}
%\begin{sidewaystable*}[htdp]
\begin{table*} 
\caption{Simulation parameters, volumes derived from this and global 
(mass averaged) reionization history results
for simulations with WMAP5 cosmology parameters. The box sizes can be directly 
inferred from the simulation names. For all simulations, the mesh consists 
of $256^3$ cells. The ionization time step for all simulations is 
$\Delta t_i = 5.75 \times 10^6$ yr. $g_{\gamma}$ is the efficiency parameter 
as explained in the text; the old efficiency paramter $f_{\gamma}$ is given 
in brackets; $V_{\rm{min}}$ is the comoving volume of the minimum size H~II 
region, ionized by the least efficient source during one time step, assuming 
the density to be the average density of the universe; $\tau_\mathrm{es}$ is 
the electron scattering optical depth calculated for each simulation.  
}

\begin{center}
\begin{tabular}{@{}l||llllllll}

& \small{53Mpc\_g8.7\_130S} & \small{163Mpc\_g8.7\_130S} & \small{53Mpc\_g8.7\_130} & \small{53Mpc\_g1.7\_8.7S} & \small{53Mpc\_g0.4\_5.3} & \small{53Mpc\_uvS\_1e9} & \small{53Mpc\_g10.4\_0} \\[2mm]\hline\\ 
$^{\rm{high}}_{\rm{mass}}$ $g_{\gamma}$ ($f_{\gamma}$)
& 8.7 (10)            & 8.7 (10)     & 8.7  (10)      & 1.7 (2)         &  0.4 (0.4)       & variable $\rightarrow$Fig.\ref{ggamma_evol} & 10.4 (12) \\[2mm]
$^{\rm{low}}_{\rm{mass}}$ $g_{\gamma}$ ($f_{\gamma}$)
& 130  (150)          & 130  (150)   & 130  (150)     & 8.7   (10)      &  5.3 (6)     & 0 (0)                 & 0 (0)   \\[2mm]
suppression                 
& yes               & yes            & no             & yes             & no           & n/a                   & n/a    \\[2mm]\hline\\
$V_{\rm{cell}}$/Mpc$^3$
& 0.0088            & 0.2575         & 0.0088         & 0.0088          & 0.0088       & 0.0088                & 0.0088  \\[2mm]
$V_{\rm{min}}$/Mpc$^3$
& 0.1361            & 0.1361         & 0.1361         & 0.0136          & 0.0084       & $>$0.1361             & 0.1627  \\[2mm]\hline\\
$z_{10\%}$                 
& 13.6              & 13.3           & 15.8           & 10.1            & 11.7         & 13.7                  &  10.5  \\[2mm]
$z_{30\%}$                  
& 10.6              & 10.4           & 14.6           & 8.5             & 10.3         & 10.6                  &  9.6  \\[2mm]
$z_{50\%}$                  
& 9.7               & 9.4            & 14.0           & 7.7             & 9.7          & 9.7                   &  9.1  \\[2mm]
$z_{70\%}$                   
& 9.2               & 8.9            & 13.6           & 7.3             & 9.3          & 9.3                   &  8.8  \\[2mm]
$z_{99\%}=z_{\rm{ov}}$                  
& 8.6               & 8.3            & 13.0           & 6.7             & 8.6          & 8.5                   &  8.3  \\[2mm]
$\tau_\mathrm{es}$                  
& 0.083             & 0.080          & 0.13           & 0.058           & 0.078        & 0.084                 &   0.071 \\[2mm]

\end{tabular}
\end{center}
\label{table_summary}
%\end{sidewaystable*}[htdp]
\end{table*}
\begin{figure} 
\begin{center}
 \includegraphics[width=8cm]{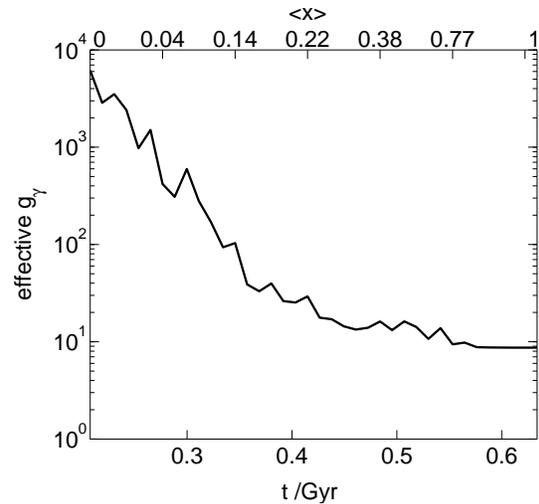}
\caption{Evolution of the effective efficiency factor $g_{\gamma}$ for the 
simulation with imposed photon production history
(53Mpc\_uvS\_1e9) as a function of time and global ionization fraction.
\label{ggamma_evol}
}
\end{center}
\end{figure} 
\begin{figure*} 
\begin{center}
 \includegraphics[width=16cm]{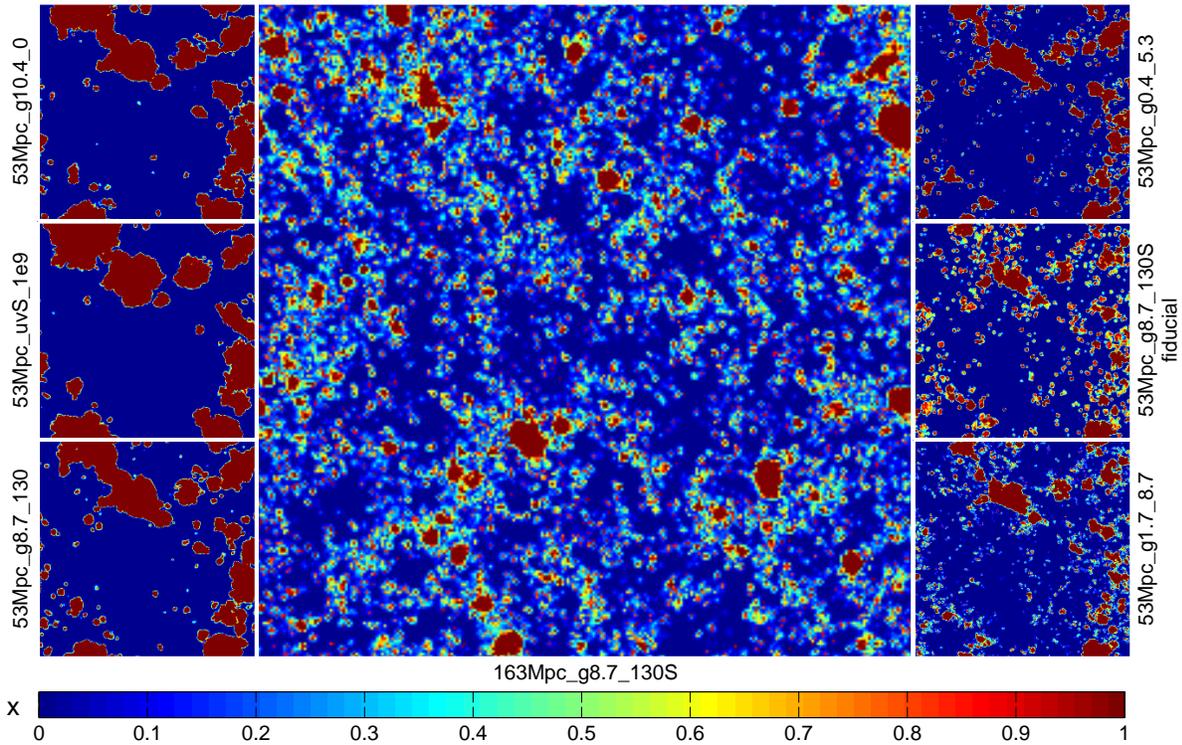}
\caption{Ionization maps (ionization fraction according to colour bar) of all included
  simulations at 30\%  global ionized fraction: from top left to bottom right:
  53Mpc\_g10.4\_0,
  53Mpc\_uvS\_1e9,
  53Mpc\_g8.7\_130,
  163Mpc\_g8.7\_130S, 
  53Mpc\_g0.4\_5.3, 
  53Mpc\_g8.7\_130S and
  53Mpc\_g1.7\_8.7S.
  Each panel is for a slice which is one cell thick ($\sim0.64$ Mpc and $\sim 0.21$ 
  Mpc, respectively for the 163 Mpc and 53 Mpc simulations). 
\label{supp_comp}
}
\end{center}
\end{figure*}

Our simulation methodology has been previously described in detail
\citep{2006MNRAS.369.1625I, methodpaper,2007MNRAS.376..534I}. Here, we
will briefly summarize the underlying N-body simulations that were
performed and the set of radiative transfer simulations that we analyze.
On the scales of interest to us here, the intergalactic medium and
dark matter followed each other as cosmic structure arose in the $\Lambda$CDM 
universe during this epoch.  Even during reionization,
 since ionization fronts which reionized the IGM moved supersonically,
the back reaction of the gas due to mass motions related to pressure forces 
can be neglected to first approximation \citep[][]{1987ApJ...321L.107S},  
and, hence, the radiative transfer can
 be done as a post-processing of the N-body density field. 

\subsection{N-body simulations}
As a basis for our radiative transfer calculations, we begin with the
time-dependent density field extracted from N-body simulations of
structure formation.  We use the \textsc{CubeP$^3$M} code which was
developed from the PMFAST code \citep{2005NewA...10..393M}, see \citet{2008arXiv0806.2887I} 
for a short description of the\textsc{ CubeP$^3$M} code. It uses particle-particle interactions at sub
grid distances and a particle-mesh method for larger distances.
Here, we use the results of two simulations, performed with \textsc{CubeP$^3$M}, 
one for a volume of 163~Mpc on 
a side, another with 53~Mpc. The former  has $3072^3$ particles
and mesh size of $6144^3$ cells, while the latter has $1024^3$ particles and 
$2048^3$ cells, which imply particle masses of $5.5\times 10^6$~M$_\odot$ 
and $5.1\times 10^6$~M$_\odot$, respectively. These parameters guarantee a
minimum resolved halo mass of $10^8$~M$_\odot$ which is approximately the minimum mass 
of halos able to cool by atomic hydrogen cooling. The cosmological parameters
used were for a flat $\Lambda$CDM universe with $\Omega_m=0.27$,
$\Omega_b=0.044$, $h=0.7$, $n=0.96$, and $\sigma_8=0.8$, based on the
five year WMAP results \citep{2009ApJS..180..330K}.
%-------------------------------------------------------------------------------
\subsection{Radiative transfer runs} 
%-------------------------------------------------------------------------------
Table \ref{table_summary} gives an overview of the seven different
radiative transfer runs that we analyze in this paper.  These are a
sub-set of a larger suite of simulations, to be presented in a
follow-up paper (Iliev et al., in prep.). This sub-set was chosen as
the minimum one needed to illustrate the points we want to make in
this work.  All the radiative transfer simulations were performed
using the \textsc{C$^2$-Ray} method \citep{methodpaper} on a uniform
rectilinear grid containing $256^3$ grid cells. The density is
assigned to the mesh by smoothing the dark matter particle
distribution from the underlying N-body simulation using an SPH kernel
function: each N-body particle is assigned a compact, spherical
smoothing kernel whose width is adjusted so as to encompass its 32
nearest-neighbors.  Particle mass is then assigned to the cells of our
radiative transfer grid by integrating each kernel function over the
volume of each cell it overlaps.   To convert what is a the IGM
  dark matter density into a baryon density, we assume that in the IGM
  the gas distribution follows the dark matter. This is valid on the
  scales of the radiative transfer cells (0.2 or 0.6 comoving Mpc) as
  at the mean density of the IGM they are much larger than the local
  Jeans length.

 There are physical effects below our resolution limit which
  influence reionization. These are small scale density variations
  (`clumping') and the presence of unresolved absorbers (such as
  mini-halos and the structures of unknown origin which are observed
  as Lyman Limit Systems at lower redshifts). All of these will slow
  down the reionization process as they increase the number of photons
  absorbed. In the simulations presented here we do not consider these
  effects, but see e.g.\ \citet{2006MNRAS.366..689C,
    2007MNRAS.377.1043M, 2010arXiv1003.6132A, 2010arXiv1008.0003C} for
  studies about the effect of different types of unresolved absorbers
  and e.g.\ \citet{2006MNRAS.369.1625I, 2007MNRAS.377.1043M,
    2008MNRAS.384..863I} for compartative studies of the effect of
  clumping.

Simulations are labeled with the parameter $g_\gamma$,
which is an efficiency factor for the ionizing photon production of
halos per source halo baryon per unit time. Each halo of mass $M$ is assigned a luminosity 
\beq
\dot{N}_\gamma=g_\gamma\frac{M\Omega_b}{10 \mu \Omega_0 m_p}\,, 
\eeq
where $\dot{N}_\gamma$ is the number of ionizing photons emitted per
Myr, $M$ is the halo mass, and $m_p$ is the proton mass.  Halos are
assigned different luminosities according to whether their mass is
above (``large sources'') or below (``small sources'') $10^9 M_\odot$
(but above $10^8 M_\odot$).  For example, 53Mpc\_g8.7\_130S indicates
that large sources have an efficiency $g_\gamma=8.7$, while small
sources have an efficiency $g_\gamma=130$, and the symbol ``S'' means
that the small sources are suppressed in regions where the ionization
fraction is higher than 10\%. 

In previous simulations performed with \textsc{C$^2$-Ray}, the source efficiencies
were characterized by  $f_\gamma$, the number of ionizing 
photons released per source halo baryon per
star-forming episode (i.e. per simulation time-step for updating
the source halo catalogue from the N-body results). The relation between $f_\gamma$ 
and $g_\gamma$ is given by
\beq
 g_\gamma$=$f_\gamma\left(\frac{10 \;\mathrm{Myr}}{\Delta t}\right)\,
\eeq
where ${\Delta t}$ is the time between two snapshots from the N-body
simulation. For example \citet{2008MNRAS.384..863I} considered a simulation
called f250.  In the new naming scheme this would be called g112. 
The reason for switching to a new
naming scheme is that the previous scheme hid the dependence on
the size of the time step $\Delta t$, since $f_{\gamma}$ ionizing photons
were released over a time $\Delta t$ per baryon for all the baryons
inside source halos when that step began.  This made it more difficult
to compare simulations involving different time-steps, since
the results depend on BOTH $f_{\gamma}$ AND $\Delta t$, while the instantaneous
luminosities of source halos depend only upon their ratio $f_{\gamma}/\Delta t$, 
not $f_{\gamma}$ alone.

The suite of simulations presented in Table~1 allows us to see how the
morphology and characteristic scales of reionization depend upon
various important numerical and physical parameters which are not yet
well understood. The simulation 53Mpc\_g8.7\_130S is our standard case
for this paper. We refer to this as the fiducial simulation. 
It produces an electron scattering optical depth
consistent with the 1--$\sigma$ range allowed by the seven year WMAP
results, $\tau_\mathrm{es}=0.088 \pm 0.015$
\citep{2010arXiv1001.4538K}. To test the effect of weaker sources, and
thus more extended reionization we also present 53Mpc\_g1.7\_8.7,
which ends considerably later and has an optical depth consistent with
the seven year WMAP results when considering the 2--$\sigma$ 
range (and assuming a gaussian error distribution) for
$\tau_\mathrm{es}$. These two simulations are used to introduce the
different analysis methods in section \ref{ana_method_intro}.

One of the physical effects which may be present during reionization and
which we study in this paper,
is source suppression due to Jeans mass filtering, 
in which ionizing radiation from sources hosted
by halos with a mass below some threshold is suppressed when the halos
are located within ionized regions \citep[e.g.][]{1994ApJ...427...25S}. 
This concept was introduced  in 
our simulation models in \citet{2007MNRAS.376..534I}. By comparing, for 
example 53Mpc\_g8.7\_130S to 53Mpc\_g8.7\_130 (the latter without source
suppression), it is possible to isolate the effects due solely to
source suppression.  However, the simulation with no suppression will
end at a much higher redshift and therefore the halo populations are
not comparable at corresponding stages of reionization (e.g., at
$z_{50\%}$).  Hence, we also include the simulation
53Mpc\_g0.4\_5.3 which does not have suppression, but which due to the
weaker source luminosities, ends approximately at the same time as our 
fiducial simulation 53Mpc\_g8.7\_130S. This way the different
reionization stages (except the earliest ones) occur at similar times, 
and thus these two simulations have
similar halo populations at the different stages of reionization. 

For the 54~Mpc simulation volume at $z \sim 13.6$, there are roughly $330$ cells 
containing source halos more massive than $10^9 \rm M_{\odot}$ (this corresponds to roughly $2 \times 10^{-3}$ \% of all cells). Additionally, 
there are about $38\,000$ cells ($2.5 \times 10^{-1} \%$) containing low mass source halos between 
$10^8 \rm M_{\odot}$ and $10^9 \rm M_{\odot}$. However, for the fiducial 
simulation for example, roughly 88\% of these low mass source halos are 
suppressed. For the large simulation volume, these numbers (also at roughly 
10\% global ionization fraction, i.e. $z\sim 13.2$) are: $12\, 000$ cells 
containing massive halos ($7.8 \times 10^{-2} \%$ of all cells) and $750 \,000$ cells ($ 4.8 \%$ of all cells) containing low mass halos, of 
which roughly 86\% are suppressed. At overlap, 
the small (large) simulation volume has about $17\,000$ ($440\,000$) cells 
containing massive source halos, which corresponds to 0.1 \% and 2.8 \% of all cells, respecively. The number for low mass halos are $280\,000$ ($3\,600\,000$) cells, or 1.8 \% and 13 \% of all cells,  almost all are completely suppressed.

Throughout this study, we will make comparisons like indicated above for the case of source suppression, in order to
see how physical effects manifest themselves and to find which
quantitative measurements best discriminate among different
reionization scenarios. Instead of comparing the simulations at equal redshifts, 
we do the comparison at equal mass averaged (global) ionization fraction, $\left< x \right>$. 
Table~1 lists the redshifts at which the global 
ionization fraction (i.e. mass-weighted average, 
unless otherwise stated) for each simulation are  
$\left< x \right> \sim 0.1,0.3,0.5,0.7,0.99$. 
The epoch at which the H II regions globally "overlap", $z_{\rm ov}$,
will, by convention, be taken here to be the redshift at which 
$\left<x \right> = 0.99$, although the value of $z_{\rm ov}$ 
which results is not very sensitive to
this particular choice as long as $\left<x \right>$ is close to unity.

Beside the above mentioned simulations, there
are three more simulations included in this study: the simulation
labeled 53Mpc\_uvS\_1e9 has the same (imposed) global photon production
history as our fiducial simulation, but only halos more massive than
$10^9 M_{\odot}$ are allowed to host luminous sources. This results in
a  $g_{\gamma}$ that is time dependent. Its evolution is plotted in  Fig.
\ref{ggamma_evol} as a function of time and mass averaged ionization
fraction $\left< x \right>$. 

As a second simulation with only high 
mass sources, we include simulation 53Mpc\_g10.4\_0. Unlike 
53Mpc\_uvS\_1e9 it has a constant mass to light ratio which is chosen 
so that reionization ends roughly at the same time as in our 
fiducial simulation. This yields later $z_{10\%}-z_{70\%}$ and
thus a lower value of $\tau_{\rm es}$.
Note that the efficiency of sources in high mass halos had to be 
boosted only by a factor $1.2$. This means that in our fiducial simulation, 
sources in low mass halos only contribute about 17\% to the ionizing photon budget. 

Simulation 163Mpc\_g8.7\_130S has the
same physical parameters and the same mass resolution and, hence, halo mass range as our fiducial
simulation, but the simulation box volume is about 30 times bigger. Therefore,
it is capable of catching structure on larger scales. On the other hand,
the resolution in the radiative transfer simulation is
worse than in the small box simulation since the number of cells in both simulations is
256 per side. This simulation is included to check for
cosmic-variance effects and to test the effect of resolution on our
investigation methods. 

Snapshots of the simulations at the 30\% global (mass averaged) ionized fraction
are shown in Figure \ref{supp_comp}. The slices are to the same comoving physical
scale to make it more easy to see the morphological and topological
differences between the models which we will discuss in detail below.

Table \ref{table_summary} also lists values for the smallest possible
H~II region $V_{\rm{min}}$ which could be formed during a single
radiative transfer time step $\Delta t_i$ if the surrounding IGM has the average
density of the Universe and recombinations can be neglected. This number
is likely to be an overestimate as recombinations and density peaks will
reduce it. However, it is a useful number to compare the resolution of
various radiative transfer simulations with. The number of emitted ionizing 
photons from the smallest halos of mass $M_{\rm{min}}$ is given by
\beq
N_{\rm{min}}=\dot{N}_{\gamma}\Delta t_i = \frac{g_{\gamma}}{10}
\frac{M_{min} \Omega_b}{\mu m_p \Omega_0} \frac{\Delta t_i}{Myr}\,, 
\eeq
which then gives a minimum volume of

\begin{eqnarray}
\nonumber
V_{\rm min}&=&\frac{N_{\rm min}}{n_H}
=\frac{g_{\gamma}M_{\rm min}}{\Omega_0\rho_{\rm
    crit,0}}\frac{\Delta t_i}{10 \rm{Myr}}\\
&\simeq& 0.0016 {\rm Mpc}^3\left(\frac{M_{\rm min}}{
    10^8 M_{\odot}}\right) g_\gamma\,,   
\label{size_est}
\end{eqnarray} 
where we have used the radiative transfer time step $\Delta t_i =
5.75 \times 10^6$ yr and $n_H=\frac{\rho_{\rm{crit,0}}}{\mu m_p}\Omega_b$ 
is the hydrogen number density. $V_{\rm min}$ can be compared to the cell-sizes 
of the simulations which are also listed in Table \ref{table_summary}.
%-------------------------------------------------------------------------------
%-------------------------------------------------------------------------------
%-------------------------------------------------------------------------------
%-------------------------------------------------------------------------------
%
% METHODS-----------------------------------------------------------------------
\section{Introducing the analysis methods} 
\label{ana_method_intro}                   
In this section we introduce our analysis methods by means
of two simulations that differ only in the mass-to-light ratio of the 
halos, 53Mpc\_g8.7\_130S and 53Mpc\_g1.7\_8.7S. Whenever we refer  
to ``the simulations'' in this section, we mean these two. 
The focus in this section is on the ability of our analysis methods to
discriminate between the two simulations. The results of all methods together 
can be seen as a characterization of the morphology of the ionization fraction. 

\subsection{Size distribution}
One of the most basic measures of reionization is the size
distribution of H~II regions. However, as will become clear below, ``size of an H~II region'' 
is a quantity which can be defined in different ways. Under the assumption that most
of the volume is either highly-ionized or highly-neutral, H~II regions
can be considered to be topologically connected volumes of space.  We
previously used a friends-of-friends (FOF) method \citep{2006MNRAS.369.1625I}
to identify such regions, using the condition  $x>0.5$ 
for a cell to be considered ionized. In contrast to this measure of the volume of connected ionized space, \citet{2007ApJ...654...12Z}
used
a different method, introduced as ``the bubble probability distribution''.
For reasons we explain below, we refer to this method as 
the ``spherical average'' method. We now describe these two methods in more detail.                                 
\subsubsection{Friends-of-friends method}
Our first method for identifying the size distribution of H~II regions
relies on a literal definition of ``H~II region'': a connected region in
which hydrogen is mostly ionized. For grid data, the
obvious way to identify such a connected region is to 
use a ``friends-of-friends'' (FOF) approach, in which two neighbouring
cells are considered friends if they both fulfill the same condition.
Cells are grouped into distinct regions according to whether they are
linked together in an extended network of mutual friends. The
algorithm we use to group cells together is the equivalence class
method, described in \citet{Numerical-Recipes}. Unless otherwise specified,
we use $x>0.5$ for a cell to be considered ionized, and $x \leq 0.5$ for a
cell to be considered neutral, so that every point in the simulation
box is either in an H~I or an H~II region. Our method was first described in
\citet{2006MNRAS.369.1625I}. In contrast to all other size measures presented below,
the FOF does not care about how contorted an H~II region is. Therefore, the sizes 
of H~II regions found by the FOF method and the sizes of H~II regions found by other methods 
which will be introduces below, give complementary information about the
morphology of the ionization fraction field. 

The FOF method has been used extensively for halo finding
in cosmological N-body simulations \citep{1985ApJ...292..371D}.  Our
implementation is more straightforward, since each cell always has only
6 direct neighbours, the identities of which are known in advance, as
opposed to particle data, in which it is necessary to perform costly
searches to identify the groups.  Another significant difference
between the two methods is the role played by free parameters.  In
the halo finding FOF method, the free parameter is the linking
length, which is the distance within which two particles are
considered to be friends.  In the region finding method, the free
parameter is the threshold, $x_{\rm{th}}$, for a cell to be considered
ionized or neutral.

\begin{figure}
\begin{center}
  \includegraphics[width=8cm]{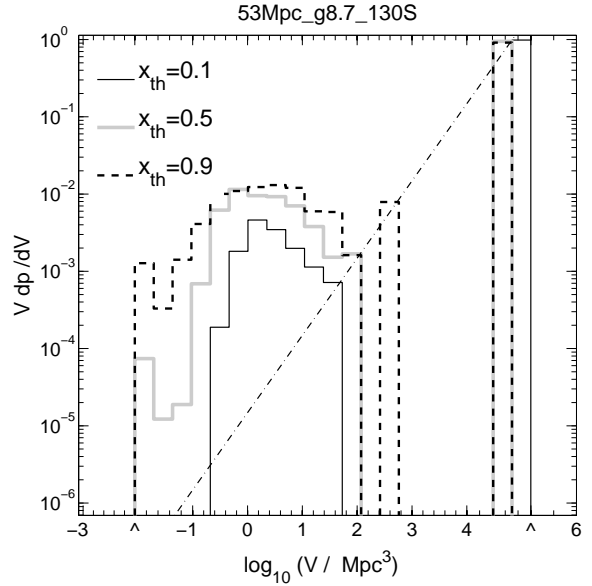}
  \caption{Effect of varying the threshold x$_{\rm{th}}$ 
  for the FOF method for the 53Mpc\_g8.7\_130 simulation at $z=9.7$, when the global 
  ionized fraction $\left< x \right>$ is about 50 percent. The results for 
three different threshold limits are shown, as indicated in the figure. 
The thin dot-dashed black line shows at every volume $V$, how much a single 
region with this volume would contribute. Indicated by \texttt{\^} on 
the abscissa are the cell- and box- volume. 
\label{fof_thresh}
}
\end{center}
\end{figure}

As seen in Fig. \ref{fof_thresh} we test the effect of varying $x_{\rm{th}}$ (using our 
fiducial simulation 53Mpc\_g8.7\_130S at 50\% global ionization fraction),  where we used three 
different values for $x_{\rm{th}}= 0.1,0.5$ and 0.9. The thin 
dot-dashed black line shows at every volume $V$, how much a single region with this
volume would contribute.  This means that each bin cannot be 
filled less (i.e. or the bin must be empty) than to the point where 
this line crosses the lower limit of each volume bin. 

As can be inferred 
from Table \ref{table_summary}, the minimum volume ionized by a 
source in a single radiative transfer time step for our fiducial 
simulation is about an order of magnitude bigger than the cell 
size. This means that at the end of the radiative transfer time step 
at which the source turned on, the cell hosting the halo will be 
completely ionized and surrounded by partly ionized cells. 
If regions are connected through such partly-ionized border cells, 
it strongly depends on  $x_{\rm{th}}$ if the regions count 
as two disconnected or as one connected region. 
In every analysis method that depends on a threshold value this effect is bigger
if the partly ionized borders of ionization regions are comparable to the regions' size. 
Nevertheless, as can be 
seen in Fig. \ref{fof_thresh}, the qualitative 
picture remains unchanged, with a few few-cell regions, a substantial 
contribution of regions of intermediate size, and the main contribution 
coming from a single large region comparable in size to the simulation box.
The absence of single-cell regions for $x_{\rm{th}}=0.1$ can be explained 
by looking at the minimum number of photons from a single source released during one time step. 
Sufficient photons are produced to ionize the cells surrounding the source cell
more than 10\%, even if the density of the cells is nine times 
the average density of the universe, see Eq. \ref{size_est}. 

Below, we convert the volume bins into equivalent radius bins,
 $R_{\rm equi,FOF}=[3/(4 \pi) V] ^{(1/3)}$ which corresponds 
to the radius $R_{\rm equi,FOF}$ of a spherical region with the 
same volume. We want to stress that this does not mean that the H~II regions are spherical. 
We convert to an equivalent radius only to allow a more 
direct comparison to the spherical average method and the power spectrum which are 
described in the following sections. We normalize to the 
total volume and not to ionized volume, to allow for a more 
direct comparison to the power spectrum which is normalized in the same way. 

\begin{figure}
\begin{center}
  \includegraphics[width=8.9cm]{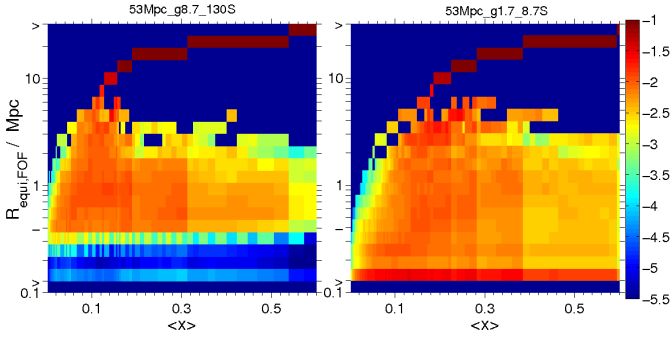}
\caption{Size distributions using the friends-of-friends
method for simulation 53Mpc\_g8.7\_130S  (53Mpc\_g1.7\_8.7S) in the left (right) 
panel as a function of global ionization fraction $\left< x \right>$. $V dp/dV$ 
is colour coded according to the colour bar. The equivalent radii corresponding 
to the cell and box- volume of the 53 Mpc box, $R_{\rm{equi,FOF}}= 0.13$ Mpc and
 $32.8$ Mpc are marked by \texttt{>}, respectively. Additionally, the cell size 
for the 163 Mpc simulations, $R_{\rm{equi,FOF}}= 0.4$ Mpc see section 
\ref{numericals}, is indicated by \texttt{-}. 
\label{fof_comp}
}
\end{center}
\end{figure}

In order to show the time-evolution of the FOF size distribution of ionized 
regions for a simulation in a single plot, we choose a fixed threshold value 
($x_{\rm th}=0.5$) and color code the contribution $V dp/dV$. The color coding 
makes it possible to show a histogram (like Fig. \ref{fof_thresh}) in a single 
line or column. Each individual column of Fig. \ref{fof_comp} (left panel) is a 
histogram as Fig. \ref{fof_thresh} at a different global ionization fraction. 
Fig. \ref{fof_comp} thus shows the evolution of the size distribution with 
global ionization fraction. The box- and the cell size are marked 
with \texttt{>} on the ordinate. By construction the FOF method will not result in 
sizes larger than the former and smaller than the latter. As we will see later, 
H~II regions start to merge very early on in the course of reionization which 
results in shapes far from spherical and a complex topology already at global 
ionization fractions $\left< x\right> \sim 0.2$. The concept of distinct H~II 
regions and their sizes quickly becomes meaningless. Therefore, all size 
distribution estimates are only shown up to a global ionization 
fraction $\left< x\right> \sim 0.6$. 
  
Three things catch the eye when analyzing the size evolution in Fig.\ref{fof_comp}: 

(1) Already at $\left< x \right> \sim 0.15 $, the distribution for both 
simulations is not continuous, but shows a gap. Most of the ionized 
volume is contained within one region of a size falling into a size-bin 
which is separated from the rest. This is an inherent property of the FOF method, 
where regions are grouped together as soon as they touch and 
local H~II region percolation occurs quite early in the evolution. If the H~II regions reach 
a certain size, which depends on the clustering of the sources 
and on their efficiency, they will percolate and form bigger H~II regions. 
As those smaller H~II regions grow and merge into the larger one, both 
their numbers and the fraction of the
ionized volume that they occupy decrease. A doubling of the volume (merging 
of two bubbles with the same size) thus corresponds to a jump over one effective radius bin. 
The largest H~II region grows
through mergers with smaller H~II regions, as well as due 
to sources within the region, and approaches the box-size by $\left< x \right> \sim 0.6$.

(2) The distribution of the smaller scales (i.e. everything except 
the one big region) is much flatter for the 
simulation 53Mpc\_g1.7\_8.7S. This simulation has more single- 
or few-cell-sized regions than our fiducial simulation 53\_Mpc\_g8.7\_130S. 

(3) While the size-bin which contributes most to the global ionization 
fraction is increasing with $\left< x \right>$, the shape of the 
distribution of the rest does not change much. However, its total contribution 
to the global ionization fraction decreases with $\left< x \right>$: At the same rate 
that small H~II regions grow bigger and merge into even bigger 
H~II regions, new small ones are ``born'' but contribute in the course of 
reionization less and less to the global ionized-fraction. 

The FOF-method applied to an ionization field in a finite simulation box can 
only sample the true underlying size distribution function (as it would be in 
an infinite simulation box) up to a lower limit. This is indicated by the thin 
dot-dashed black line in Fig.\ref{fof_thresh}. This lower limit depends on the 
size of the simulation box. The gap mentioned above will be smaller the better 
the sampling, that is the bigger the simulation box.    

\subsubsection{Spherical Average method}
The spherical average method (SPA) was described by \cite{2007ApJ...654...12Z}; it can be easily used for comparisons with analytical models. 
It is based on constructing spheres around every cell in the computational volume,
averaging the ionization fraction inside these spheres and finding the largest such 
sphere for which the average ionization fraction is greater than a certain threshold $x_{\rm th}$. 
We chose $x_{\rm th}=0.9$. Because of this, we call it the \textit{spherical average} 
method (SPA). It yields a smoother 
distribution of H~II region sizes than the one obtained by the FOF method. It does 
not measure the size of a connected ionized space but instead it is a measure of 
the scales of spherical bubbles which would cover the ionized space.  

\begin{figure}
\begin{center}
  \includegraphics[width=8.9cm]{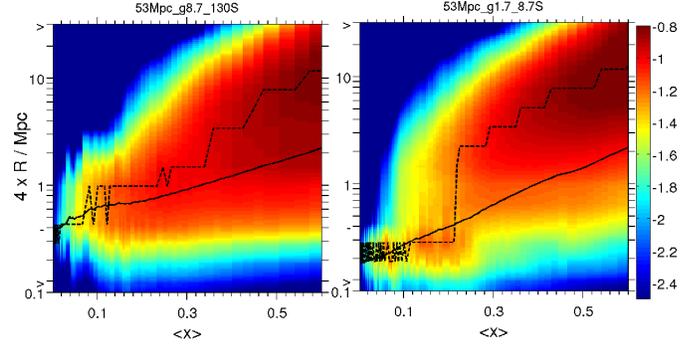}
\caption{Size distributions using the spherical average method as explained in 
the text, $\log_{10} R dP/dR$ colour coded according to colour bar; left (right) 
panel simulation 53Mpc\_g8.7\_130S (53Mpc\_g1.7\_8.7S) as a function of global 
ionization fraction $\left< x \right>$. The peaks of the distribution at every 
global ionization fraction are indicated by the dashed line. The black solid 
line shows the 3VA size measure, as explained in section \ref{3VAmethod}.
\label{spa_comp}
}
\end{center}
\end{figure}

Motivated by the analysis given in Appendix \ref{app_spa}, we multiply the 
radius found by the spherical average method by a factor of 4, $s=4 \times R$. 
We call this the scale of the spherical average method.   
In Fig. \ref{spa_comp}  we plot the spherical average distribution in the same way 
as the FOF distribution in Fig. \ref{fof_comp}. 
$R dP/dR$ is normalized to the whole volume.  
We define $R_{\rm max}(\left< x \right>)$ to be the position of the maximum 
of $R dP/dR$ at every $\left< x \right>$. We see that  $R_{\rm max}$ is 
increasing with $\left< x \right>$ which means an increase 
in the average bubble size with global ionization fraction, as expected. 
Although initially smaller, it can be seen that the average scale of 
simulation 53Mpc\_g1.7\_8.7S is growing faster with respect to 
$\left< x \right>$ than the average scale of simulation 53Mpc\_g8.7\_130S. 
From Fig. \ref{spa_comp}
it can be further seen that the distribution of the
model with the smaller efficiencies shows initially (at $\left< x \right> \lesssim 0.1$) a wider 
distribution of bubble sizes  with a peak at smaller 
scales: The smallest H~II regions are smaller than in the fiducial model, 
but the contribution from small regions to total ionized fraction is smaller. 
Therefore the bigger H~II regions have to grow to a bigger size in the 
53Mpc\_g1.7\_8.7 model to reach an ionization fraction of 10\%. At higher 
average ionization fractions, the peak of the distribution is slightly 
shifted towards bigger scales with respect to the 53Mpc\_g8.7\_130S model.   

As we have already noted, much of the noticeable difference in
the FOF curves comes from a very small fraction of the volume, and a
log scale is required to see such differences in the FOF plots.  The
spherical average distributions are much smoother, and therefore offer
a less detailed, more global picture of the spatial structure of the
ionized regions. 
However, the spherical average shows much more clearly the difference in size 
of the large scale H~II regions between the two simulations. 

{Another method which is similar to the SPA is a method used in 
\cite{2007ApJ...669..663M}. We examine this method in Appendix 
\ref{MF_size_measure}. We find that it yields qualitatively the same 
results as the SPA and PS but its use in simulations with a continuous 
distribution of ionization fractions (i.e. not a binary field) creates complications.}

\subsubsection{3V/A method}
\label{3VAmethod}
As another estimate of the scale of bubbles, one could use the ratio of the 
total volume of the H~II regions and their total surface area: 
\beq 
3 \times \sum_{\rm{H~II~regions}} \rm Volume / \sum_{\rm{H~II~regions}} \rm Area.
\eeq
The volume $V$ and the surface area $A$ were calculated from the zeroth and 
first minkowski functionals, respectively: $V=V_0$ and $A=6 \times V_1$.
For a distribution of disconnected spherical bubbles, 
3V/A is the surface weighed average radius. For three dimensional bodies, one 
could say that $3 V/A$ is proportional to the surface weighted average of the 
smallest scale of each object. If the dominant structures are 2-dimensional, 
i.e. disc-like, then $3 V/A$ is three times bigger than the surface weighted 
average disc-height. If the dominant structure is 1-dimensional, i.e. bar-like, 
$3V/A$ is 1.5 times the surface weighted average bar-radius. In any case, in 
terms of surface weighted averages, it is an overestimate of the minimum scales.
As in the case of the spherical average, $3 V/A$ does not measure the sizes of topologically 
connected ionized volumes.

It can be seen (Fig. \ref{spa_comp}) that initially, for both simulations, the 
$3V/A$ scale agrees with the scale of the maxima of the spherical average. 
At about 20\% global ionization fraction, the scale of the maxima of the 
spherical average for the 53Mpc\_g1.7\_8.7S simulation is greater than the 
scale estimated by $3V/A$, indicating that while most of the volume is contained 
in larger scale bubbles, most of the surface comes from small scale bubbles: 
many very small scale structures, few very large scale structures. For the 
fiducial simulation, the offset between the scale of the maximum of the 
spherical average and the scale estimated by $3V/A$ is smaller, indicating 
that the same bubbles which contribute substantially to the total volume, 
contribute substantially to the total surface area.    
\subsection{Power spectra}
\begin{figure}
\begin{center}
  \includegraphics[width=8.9cm]{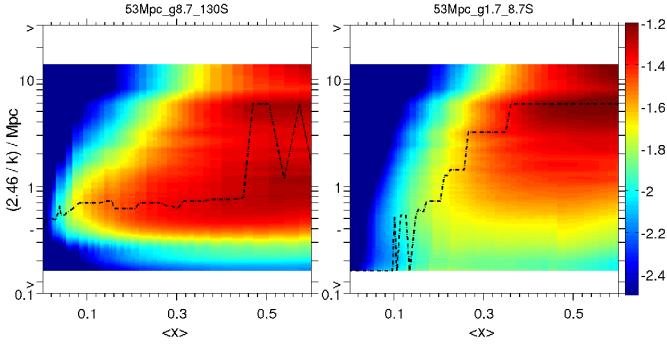}
\caption{Power spectra of ionized fraction as a function of global ionization 
fraction, $\Delta_{xx}^2(k)$ colour coded according to the colour bar; left (right) 
panel for the fiducial simulation (53Mpc\_g1.7\_8.7S). The dashed line indicates 
the peak of $\delta_{xx}^2(k)$ as a function of $\left< x \right>$. Note that 
$\Delta^2(k)$ is plotted over a multiple of inverse $k$, for details see text.
\label{dxx_comp}
}
\end{center}
\end{figure}
The typical reionization scales can be further characterized by the
appearance of features in the power spectra of the density and ionized
fraction fields, $P_{\delta\delta}$, $P_{x\delta}$, and $P_{xx}$,
where $\langle \delta_{\bf k}\delta_{\bf k'}^{*}\rangle\equiv
(2\pi)^3\delta^3({\bf k}-{\bf k'})P_{\delta\delta}(k)$, $\langle
\delta_{x\bf k}\delta_{\bf k'}^{*}\rangle\equiv 
\delta^3({\bf k}-{\bf k'})(2\pi)^3P_{x\delta}(k)$, and $\langle
\delta_{x\bf k}\delta_{x\bf k'}^{*}\rangle\equiv 
\delta^3({\bf k}-{\bf k'})(2\pi)^3P_{xx}(k)$.  Here, $\delta$ is the
overdensity of matter, while $\delta_x\equiv x-x_v$, where $x_v$ is the 
volume-weighted global average ionized fraction. Note that we do
not normalize $\delta_x$ by $x_v$.  When plotting the actual power
spectrum, we use the dimensionless power per logarithmic interval in
wavenumber, $\Delta^2(k)\equiv k^3P(k)/(2\pi^2)$. 

The power spectrum of the ionization fraction field is one component in the expansion
of the neutral hydrogen power spectrum \citep[e.g.][]{2006PhR...433..181F}. 
As pointed out in section 1, future 21--cm radio observations aim at observing this
quantity via the 21cm line of neutral hydrogen. Therefore, from all methods presented here 
to characterize the morphology of the ionization fraction field, the power spectrum of the
ionization fraction is the measure most closly related to observations.  

Shown in Figure \ref{dxx_comp} is the ionized fraction power
spectrum, $\Delta^2_{xx}(k)$, for the two simulations as a function 
of global ionization fraction on the abscissa. Instead of plotting 
$\Delta^2(k)\equiv k^3P(k)/(2\pi^2)$ against $k$, we choose to plot 
it against $2.46/k$, the reason for this will become clear later. This colour contour 
plot thus shows the evolution of $\Delta_{xx}(k)$ during reionization. For the 53Mpc\_g8.7\_130S 
model it can be seen that the power spectrum first peaks at scales of the 
order $s=2.46/k_{\rm max}\sim 0.8$ ~Mpc. We expect this peak to be associated with the
the size of ionized or neutral bubbles, for the following simple
reason. On scales smaller than the bubbles, the correlation function 
$\xi_{xx}(r_{12})=\langle (x({\bf r}_1)-x_v)(x({\bf r}_2)-x_v)\rangle$
reduces to the constant value $x_v(1-x_v)$, while on scales much
larger than the bubbles, the ionized fraction is uncorrelated, and the
correlation function should approach zero \citep{2004ApJ...608..622Z}. This
behavior for the correlation function implies that the power spectrum
$\Delta^2(k)$ should approach zero at large and small scales, with a
peak at the characteristic size of the bubbles.  

\begin{figure}
\begin{center}
  \includegraphics[width=4cm]{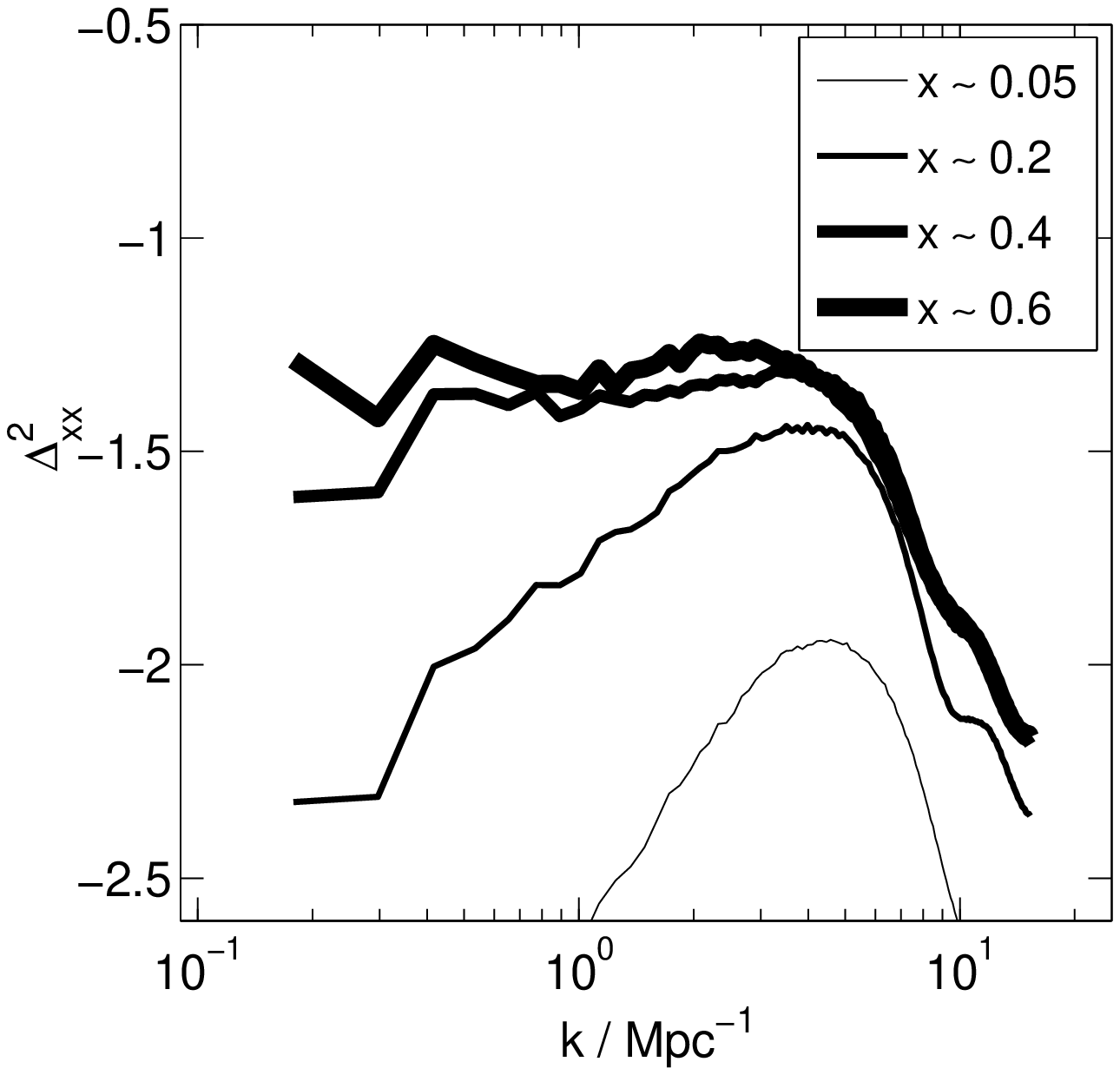}
  \includegraphics[width=4cm]{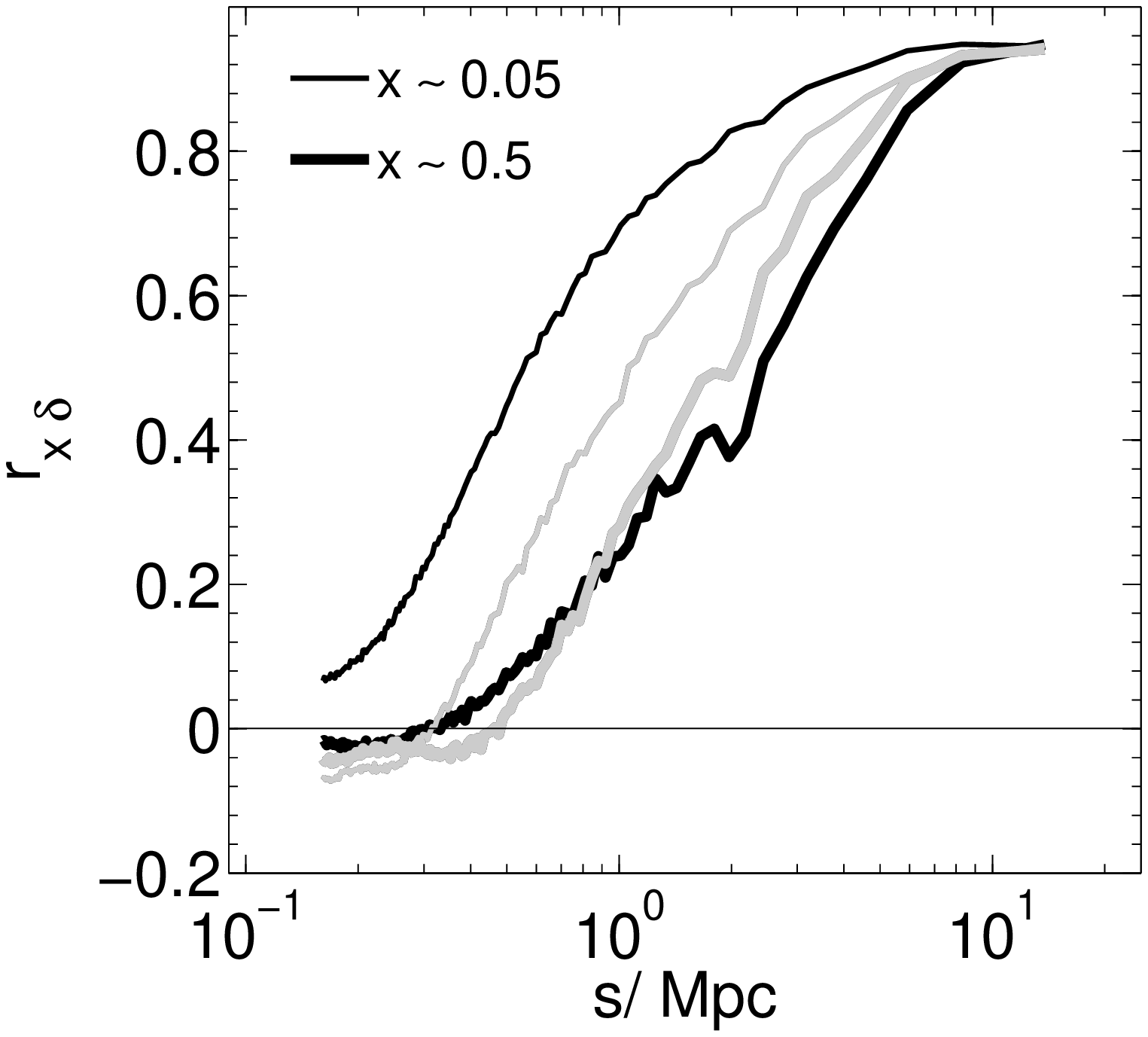}
\caption{left panel: Power spectrum of the fiducial simulation at several different global ionization fractions as indicated in the legend; right panel: Cross-correlation coefficient $r_{x \delta}$ as defined in Eq.~\ref{cross_corr_coef} 
as a function of $2.46/k$ at $\left< x \right>= 0.05$ and $0.5$ (as indicated in the figure 
(by line thickness) for the fiducial simulation (gray lines) and 53Mpc\_g1.7\_8.7S 
(black lines). Note the lower correlation between ionized fraction and density 
for the fiducial simulation on all scales at low global ionization fractions. 
\label{rcc_comp}
}
\end{center}
\end{figure}

The first peak of 
$\Delta_{\rm{xx}}^2(k)$ for a single spherical top-hat bubble of 
radius $s$ would be located at $k_{\rm{max}}\approx 2.46/s$, which is why we 
use $2.46/k_{\rm max}$ to characterize the typical radius of regions. 
Indeed, comparison of the maxima of the spherical average model and the peaks 
of the power spectra shows that 
they are approximately related by $2.5/k_{\rm max}\sim 4 R_{\rm max} $ as 
can be seen by comparing Fig. \ref{spa_comp} and Fig. \ref{dxx_comp}.
It should be noted that both, the spherical average and the power spectrum 
do not show a pronounced peak at 
all global ionization fractions, but are instead rather flat at higher global 
ionization fractions. This is in agreement with one of the trends that 
\cite{2010arXiv1003.3455Z} identify for all their tested reionization models. 
To better show this behaviour of the PS at higher global ionization fractions, 
we show the power spectrum of the fiducial simulation 
as a function of $k$ for four different global ionization fractions as a line-plot 
in the left panel of Fig. \ref{rcc_comp}. However, the other trend they identify,
the shift towards smaller $k$ with increasing lower ionization fraction is less
apparent for our fiducial simulation. Instead, the form of the power spectrum of the fiducial simulation 
suggests that 
at global ionization fraction greater than $\left< x \right> \sim 0.4 $ there are two
main populations of bubbles: One at sizes where the PS initially peaks, around 0.8~Mpc 
and the other with sizes around 6~Mpc, which is of the size of clusters of 
galaxies, i.e. the typical clustering scale of the source halos at this epoch,
not to be confused with the length scale which encompasses the
mass of the higher-mass, virialized 
galaxy clusters familiar at lower redshift. The first scale results from 
suppression of sources after initial turn on of a source in a low-mass halo. 
Growth of the H~II region is completely halted until a high-mass halo 
forms in that region. A less stringent suppression criterion would slow down the 
growth but probably not halt it entirely. 
The second scale results from merging of bubbles emerging from galaxies in 
the same cluster. These details in the size distribution are washed out in 
the spherical average distribution. However, it should be noted that a scale 
of 6~Mpc is a substantial fraction of a 53~Mpc box and therefore it is 
questionable if the sampling at this scale is high enough.     
At global ionization fractions 
larger than $\left< x \right> \sim 0.4$, it can be seen that there is 
considerable power on scales comparable to the box size. 

For the 53Mpc\_g1.7\_8.7S simulation it can be seen that there is more power 
on smaller scales but also that the power spectrum is flatter than the one of the fiducial
 model as there is no distinct peak below global 
ionization fractions of 20\%. Further it can be seen that the slope of isochromatic 
lines is greater for this model than for the fiducial one. This means that
the sizes of H~II regions in the 53Mpc\_g1.7\_8.7S are growing faster with 
respect to global ionization fraction. This was also seen in the spherical average results.
The absence of the peak at smaller scales that is present in the fiducial model, 
is due to the fact that already at a global ionization fraction of roughly 10\%, 
the contribution from sources hosted by massive halos is about the same as the 
contribution from sources in low-mass halos, while this is true at about 70\% 
global ionized fraction for our fiducial simulation. Therefore, the relative 
contribution from H~II regions produced by sources in low mass halos in isolated 
cells is smaller. H~II regions produced by sources in massive halos will grow 
continuously, explaining the flat distribution below several Mpc.   

The right panel of Fig. \ref{rcc_comp} shows the cross-correlation coefficient of
ionized fraction and density field, 
\beq
r_{x\delta}(k)\equiv
\frac{\Delta^2_{x\delta}(k)}
{\left[\Delta^2_{xx}(k)\Delta^2_{\delta\delta}(k)\right]^{1/2}}
\label{cross_corr_coef}
\eeq 
at two different $\left< x\right> \sim 0.05$ and $ 0.5$ for both 
simulations plotted against $s=2.46/k$.
When $r_{x\delta}=(-1)1$, the ionized fraction and density field are
perfectly (anti-)correlated, while $r_{x\delta}=0$ implies they are
uncorrelated. As seen from the figure, the ionized fraction and
density fields are nearly perfectly correlated on large scales, $s\gtrsim
8$~Mpc.  
It can be seen that the scale $s$ at which 
the correlation starts to decrease, is increasing with global ionization 
fraction. This is due to the fact that while the H~II regions grow, they 
also start ionizing the voids. 
At very low global ionization fractions, represented here by $\left< x\right> \sim 0.05$, 
the correlation coefficient for simulation 53Mpc\_g1.7\_8.7S is greater than 
the one for the fiducial simulation (especially at smaller $s$). This is expected because
the ionizing radiation of the sources in the simulation with lower efficiencies 
can less easily "break out" of high density regions. Additionally, less efficient 
sources  trace the high density regions better since clustered low mass 
sources are less suppressed: individually they form smaller H~II 
regions and therefore do not suppress each other. At later stages of reionization 
this difference disappears: The simulation with low source efficiencies reaches 
the same global ionization fraction as the fiducial simulation at much later 
times when massive sources are more common. Those massive sources form bigger 
HII regions which also grow into the voids. 

\subsection{Topology of Reionization} 
\label{topology}
Minkowski functionals have been used extensively in cosmology to 
characterize the topology of large scale structure 
\citep{1986ApJ...306..341G,1994A&A...288..697M,1997ApJ...482L...1S} and also the 
non-Gaussianity of the cosmic microwave background 
\citep{2009ApJS..180..330K}.  Recent work has focused on using 
Minkowski functionals as a way to characterize the morphological 
structure of reionization \citep{2006MNRAS.370.1329G,2008ApJ...675....8L}.

Both works focused on the topology of the H~I density field.
They showed the Euler Characteristic (or genus, respectively) as a function of 
neutral density  for several different times (i.e. global ionization fractions) and
concentrated on the increasing deviations from the typical curve of a gaussian 
random field. Here, we will take a complementary approach based upon the topology 
of the ionization fraction field, rather than the fluctuating neutral density 
field. Unlike the neutral density field, which takes values spread continuously 
over a very wide range, the ionized fraction field ranges only between
0 and 1, and, ideally, there would essential be only two
values possible to assign to any given point in space, either ``neutral'' values 
close to zero or ``ionized'' values close to unity. In that ideal case, the Euler 
Characteristic for the ionized fraction field would only be a function of time 
(or of the evolving globally-averaged ionized fraction) and be largely independent
 of the choice of ionization fraction threshold.

We follow the definition and notation of \citet{1996dmu..conf..281S} and 
\citet{1997ApJ...482L...1S}. 
Consider a scalar function $f({\bf x})$ defined at each point ${\bf x \in \mathbb{R}^3}$. The set 
$F_{\rm th}$ of all points ${\bf x}$ for which $f({\bf x}) > f_{\rm th}$ defines bodies 
in three dimensional space. The zeroth Minkowski functional,  $V_0(f_{\rm th})$,
is simply the volume of those bodies:
\beq
V_0(f_{\rm th})=\int_V\Theta[f_{\rm
  th}-f({\bf x})] d^3{\bf x},
\eeq
where $\Theta$ is the Heaviside step function. The next three
  Minkowski functionals are defined as surface integrals over the
  boundary of the bodies:
\begin{eqnarray}
V_1(f_{\rm th})&=&\frac{1}{6}\int_{\partial F_{\rm
    th}}({\bf x})d^2A\\
V_2(f_{\rm th})&=&\frac{1}{6\pi}\int_{\partial F_{\rm
    th}}({\bf x})
\left(\frac{1}{R_1}+\frac{1}{R_2}\right)d^2A\\
V_3(f_{\rm th})&=&\frac{1}{4\pi}\int_{\partial F_{\rm
    th}}({\bf x})
\frac{1}{R_1R_2}d^2A,
\end{eqnarray}
where $R_1$ and $R_2$ are the principal radii of curvature along the
surface $\partial F_{\rm th}$.  The first Minkowski functional is 
proportional to 
%one sixth of 
the integrated surface area. This and 
the zeroth Minkowski functional were used when calculating the size 
estimator $3V/A$. The Minkowski functional $V_3$, which
is proportional to the integral of the Gaussian curvature over the
surface, is also known as the Euler characteristic, and is equal to: \\
\noindent
$\# \rm{parts}- \# \rm{tunnels} + \# \rm{cavities}$ \\
\noindent
Applied to the ionization fraction field, two disconnected ionized cells would 
count as two parts, a ring-like ionized region constitutes a tunnel and one part 
and a neutral cell completely surrounded by ionized cells is a cavity.
Table 1 in \citet{1996dmu..conf..281S} gives an overview of different notations 
for the Minkowski functionals which only differ in constant factors.
The better known quantity genus, $g$ (number of complete cuts one can make through 
the object without dividing it into disconnected parts), is related to the Euler 
characteristic by the simple relation 
$g=1-V_3$. \footnote{This is true if one considers the Euler Characteristic of the 
volume defined by the set of points. Note that others consider the Euler 
characteristic of the surface of the set of points defining the volume. In this 
case, the Euler Characteristic $\chi$ is a factor of two greater, resulting in 
the relation to genus: $\chi = 2(1-g)$. This is consistent with the relation 
$\chi(\partial A)= \chi(A)(1+(-1)^{d-1})$, where $d$ is the dimension, $A$ a 
$d$-dimensional body and $\partial A$ its $d-1$-dimensional surface, see 
equation (18) in \citet{1994A&A...288..697M}.}
For example, a torus has $V_3=0$, since it has zero total curvature, 
and has one part and one tunnel.  A sphere, on the other hand, has
$V_3=1$, since it has one part and no tunnels, and positive total
curvature. 

\begin{figure}
\begin{center}
  \includegraphics[width=8.2cm]{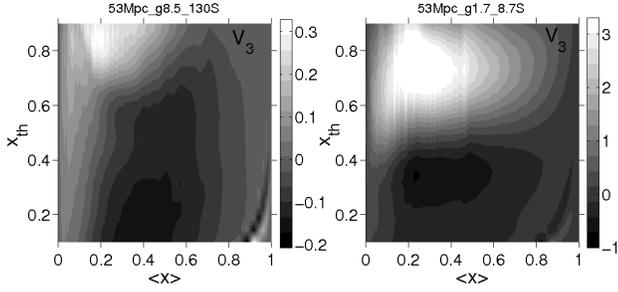}
\caption{
Euler characteristic $V_3$ as a function of global ionization 
fraction $\left< x \right> $ and threshold value $x_{\rm th}$ for the fiducial 
simulation (simulation 53Mpc\_g1.7\_8.7) in the left (right) panel. The fast 
changes in the lower right corner are due to the implementation of photons that 
would travel distances longer than the box-size, see explanation in the text. 
Note the high dependence on threshold value for simulation 38Mpc\_g1.7\_8.7. 
\label{xivx}
}
\end{center}
\end{figure}

We oversample the ionization fraction fields before calculating the Euler 
characteristic. We do this to minimize critical connections of H~I and H~II 
regions. A critical connection is for example an ionized cell which is  connected 
via an edge to another ionized cell in an otherwise neutral neighbourhood. 
Appendix \ref{minkapp} explains in more detail the problems that are involved. 
Important for the following analysis is to note that oversampling reduces 
ambiguities concerning the connectivity at a given threshold value but 
introduces higher dependences of $V_3$ on the threshold value.

In the left panel of Fig. \ref{xivx} we show the evolution of the Euler characteristic 
$V_3$ of the ionized fraction as a function of the threshold $x_{\rm  th}$ for 
our (oversampled) fiducial simulation. We choose to normalize $V_3$ by dividing by the
box size, to have an easier comparison when dealing with different box sizes. 
Between threshold values $x_{\rm th} = 0.2$ and $0.6$, the evolution of $V_3$ is 
largely independent on the actual choice of  $x_{\rm th}$: $V_3$ rises to a 
maximum value at a mean ionization fraction of about 5\%, after which $V_3$ 
decreases and gets negative before 20\% global ionization fraction is reached. 
It rises again after the ionization fraction passed 50\% but never reaches positive values again.
\begin{figure} 
\begin{center}
 \includegraphics[width=8cm]{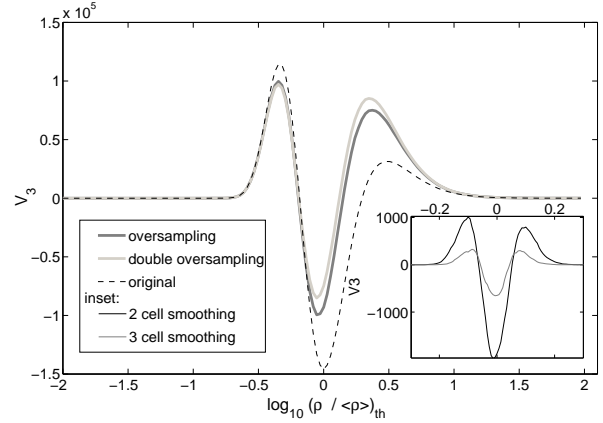}
\caption{Euler Characteristic $V_3$ of the density field as a function of 
density threshold value $\log_{10}(\rho/\left<\rho\right>$. Shown is $V_3$ 
of the original density field at redshift $z=26.1$ and  $V_3$ of oversampled 
fields as indicated in the figure. The inset shows $V_3$ of the smoothed 
field with two different smoothing width, as indicated.   
\label{mink_dens}
}
\end{center}
\end{figure}

This behaviour can be qualitatively understood by considering inside-out 
reionization in an approximately Gaussian density field.
$V_3$ for a Gaussian random field (or any monotone and steady function of it) 
as a function of threshold value, as plotted for example in figure 1 in 
\citet{1997ApJ...482L...1S}, shows in the second half a rise to positive 
values and decreases again. In Fig. \ref{mink_dens}, we show $V_3$ of the 
density field from the 53 Mpc box simulation at redshift $z \sim 26.1$ as a 
function of density-threshold value \footnote{$V_3$ of the original density 
field shows an asymmetry between isolated regions and isolated cavities, for 
details see Appendix \ref{minkapp}. A Gaussian smoothing with $\sigma \sim 3$ 
cells would be necessary to account for this. This removes all small scale 
structure, therefore $V_3$ is substantially reduced. It also removes extreme 
over- and under- densities which is why the $V_3$-curve gets narrower. Using 
sub-grid sampling enhances cells with intermediate densities and therefore 
changes the distribution away from gaussian distribution. Therefore we see 
deviations from the expected curve for a gaussian random field, which has 
$|V_3^{\rm min}|/|V_3^{\rm max}| = 1/(2 \exp(-3/2) ) \sim 2.2$, as can be 
calculated for example with equation (14) in \citet{1997ApJ...482L...1S}.}.
 The fact that $V_3$ of the ionization field does not show a rise to positive
 values followed by a decrease (see Fig. \ref{xivx} or Fig. \ref{xivx_b}), 
shows that if there exists a monotonic steady functional relation between 
density and time of ionization it does so only until a certain density: 
low-density areas (the voids in the density field) are ionized ``before 
their time'' and do not serve as positive contributions to $V_3$ in the 
ionization field. This is another feature of inside-out reionization: 
the H~II regions do eventually break out in the voids.

It can be seen that $V_3$ changes very fast at high global ionization fractions 
especially when choosing lower threshold values, see lower right corner in the 
left panel of Fig. \ref{xivx}. This is due to the way photons that would travel 
further than a box-distance are implemented in the simulations: The lost photons
 are collected and evenly distributed over all cells. Cells that only get ionized
 by those photons have small ionization fractions that only depend on their density. 

For very high values of  $x_{\rm th}$, the connectivity is reduced, therefore 
$V_3$ is greater due to the positive contribution from many more disconnected 
H~II regions. This reduced connectivity is due to many partly ionized cells 
originating partly by the relatively small photon-output per halo per ionization 
time step compared to the average number of atoms in a cell and partly by the 
oversampling that introduces additional partly ionized cells.  

It can be seen (Fig. \ref{xivx}, right panel) that the dependence of $V_3$ on 
the threshold value is more pronounced for simulation 53Mpc\_g1.7\_8.7S. From 
the FOF investigations (see Fig \ref{fof_comp}), we know that this simulation 
has many cell-size H~II regions. Also the simple estimates in Table 1 show that 
the smallest sources are not efficient enough to ionize their own cell, producing 
partly ionized cells. Therefore, a strong dependence on $x_{\rm th}$ is to be 
expected. This shows that the resolution for this choice of source efficiencies 
is not sufficient for doing topological investigations. Due to our choice of 
oversampling the data, choosing a higher threshold value ($x_{\rm th} \sim 0.5)$ 
corresponds to underestimating the connection of H~II regions. Choosing a lower 
threshold value might overestimate the connectivity. Comparing values of $V_3$ at 
higher and lower threshold values, can be used as an indication of the sufficiency 
of resolution of the simulation. In the remainder of this work, we will mostly only 
show the evolution of $V_3$ at threshold value $x_{\rm th}=0.5$ of the oversampled
 data-fields and indicate cases where $V_3$ is highly dependent on the threshold 
value. Since the extrema of $V_3$ of different simulations can vary quite a bit, 
we choose to plot the function $f(V_3)=\rm{Re} {\sqrt{V_3}}-\rm{Im}{\sqrt{V_3}}$ 
instead of just $V_3$. In the interval $[V_3^{max},V_3^{min}] 
\rightarrow [f(V_3^{\rm max}), f(V_3^{min})]$, $f$ is bijective 
(i.e. the function has an inverse function), therefore we continue to refer to it as $V_3$. 

%-------------------------------------------------------------------------------
%-------------------------------------------------------------------------------
%-------------------------------------------------------------------------------
% SECTION 4 --------------------------------------------------------------------
\section{Box-size and resolution}   
\label{numericals}                  
\begin{figure*}
\begin{center}
  \includegraphics[width=15cm]{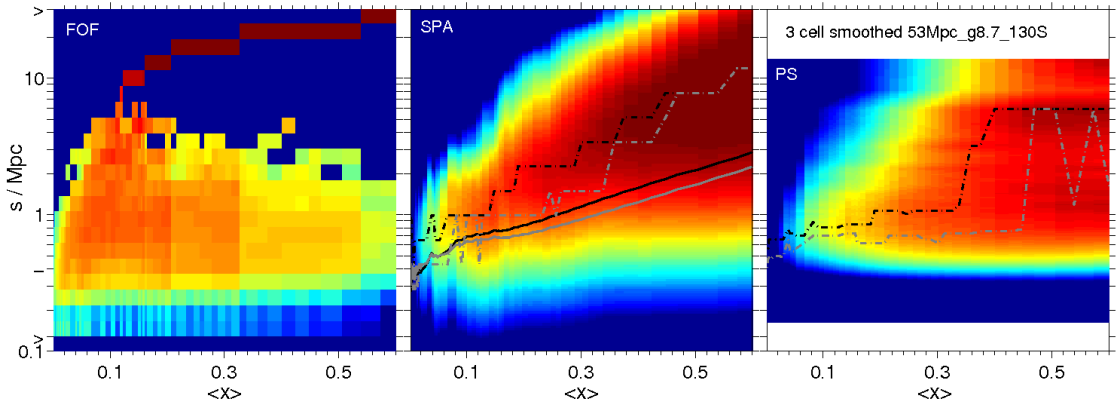}
\caption{Color plots of evolution of size distributions for the three-cell 
smoothed data of simulation 53Mpc\_g8.7\_130S. The left panel shows the FOF 
distribution (color coded as in Fig. \ref{fof_comp}). The middle panel shows 
the SPA (color coded as in Fig. \ref{spa_comp}) and its peaks (black dot-dashed line) 
together with the $3V/A$ estimate (black solid line). The right panel shows the 
PS (color coded as in Fig. \ref{dxx_comp}) and its peaks (black dot-dashed line). 
For comparison, the gray lines show the corresponding measures of the fiducial 
simulation. Indicated on the ordinates are the cell and simulation volume size 
for the 53 Mpc simulation  (\texttt{>}) and the cell size of the 163 Mpc 
simulation volume (\texttt{-}).   
\label{fof_comp2}
}
\end{center}
\end{figure*}
In this section, we investigate the effect of simulation volume size on the
simulation and the effect of resolution on our analysis methods. We compare our 
fiducial simulation to a simulation with the same source properties but in a 
bigger volume, 163Mpc\_g8.7\_130S. Before we do so, we analyze a smoothed version 
of the data of our fiducial simulation to test the effect of resolution on our 
analysis methods. We replace the ionization fraction data in each cell with the 
average over a three cell width volume centered on the cell in question, i.e. an 
average over 27 cells. We will refer to this as three-cell smoothing. This 
results in a resolution similar to the one in the 163 Mpc simulation. It should
 be kept in mind that smoothing over three cells does not remove all structure 
smaller than three cells.

In Sect. 3 we introduced three measures of size distribution and one estimate 
for the average bubble size. In Fig. \ref{fof_comp2} we plot all four measures 
as a function of global ionization fraction for the smoothed version of 
53Mpc\_g8.7\_130S. We also show curves of the peaks of the spherical average 
and the power spectrum. As a reference, these same curves (including the 
$3V/A$ estimator) for the fiducial simulation are included as gray lines. We first 
concentrate on the $3V/A$ estimate. As can be seen in the middle panel of 
Fig.\ref{fof_comp2} (comparing the two solid lines), above a global ionization 
fraction $\left< x \right> \sim 0.1$, the smoothed version yields larger values 
for $3V/A$. However, the difference is never greater than 20\%. 

The effect on the spherical average distribution (same panel) is mainly a 
reduction of contribution from scales below 0.5~Mpc. Additionally, the peak 
of the spherical average (compare dot-dashed lines in the same panel) is 
slightly shifted towards larger scales for the smoothed data.   

A somewhat contrary effect can be seen in the FOF size distribution: The 
contribution from scales below $s \sim 0.3$~Mpc is enhanced in the smoothed 
data. This can be explained as follows. If a larger ionized structure is 
elongated (i.e. no structure in 2 dimensions) and very inhomogeneous in its 
ionization fraction, then 3-dimensional smoothing would break up the 
structure in smaller parts.

In the right panel of Fig. \ref{fof_comp2} we show the power spectrum of the 
three-cell smoothed data, its maximum curve and the maxima from the fiducial 
simulation without smoothing. It can be seen, similar to the spherical average 
distribution, that power on scales below $s \sim 0.5$ is removed. Also, it 
can be seen, that at higher global ionization fraction, the distinct peak at 
scales around 0.8~Mpc diminishes while the peak at $s \sim 6$~Mpc is as 
pronounced as in the unsmoothed data.  

Since smoothing reduces the small scales and therefore reduces the critical 
(vertex/edge) H~II/H~I region-connections, oversampling the smoothed data does 
not change $V_3$ more than 10\%. It can be seen that the form of the evolution 
of $V_3$ for the fiducial simulation stays roughly the same even for an 
eleven-cell smoothing of he data. However, since the smaller scales are 
smoothed out, the total amplitude of $V_3$ is reduced.

\begin{figure}
\begin{center}
  \includegraphics[width=8cm]{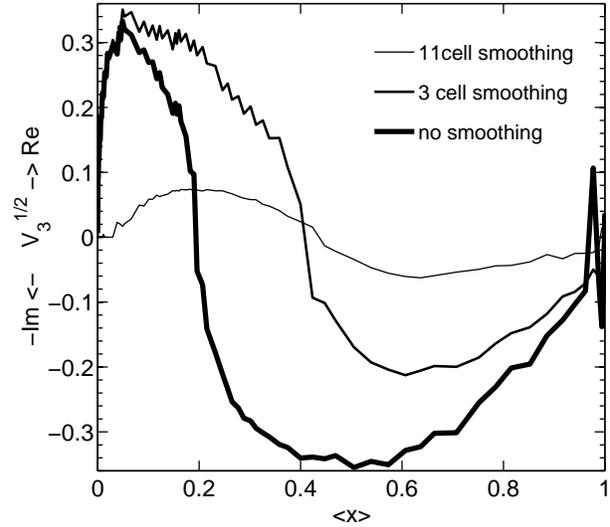}
\caption{Evolution of $V_3$ for smoothed versions (degree of smoothing as 
indicated in the figure by line-thickness) of the fiducial simulation at 
threshold value $x_{\rm th}=0.5$. Note the similarity of the behavior of 
the curves for different smoothing lengths. 
\label{mink1}
}
\end{center}
\end{figure}
%
%
%-------------------------------------------------------------------------------------------------------------------------------------
\begin{figure*}
\begin{center}
  \includegraphics[width=15cm]{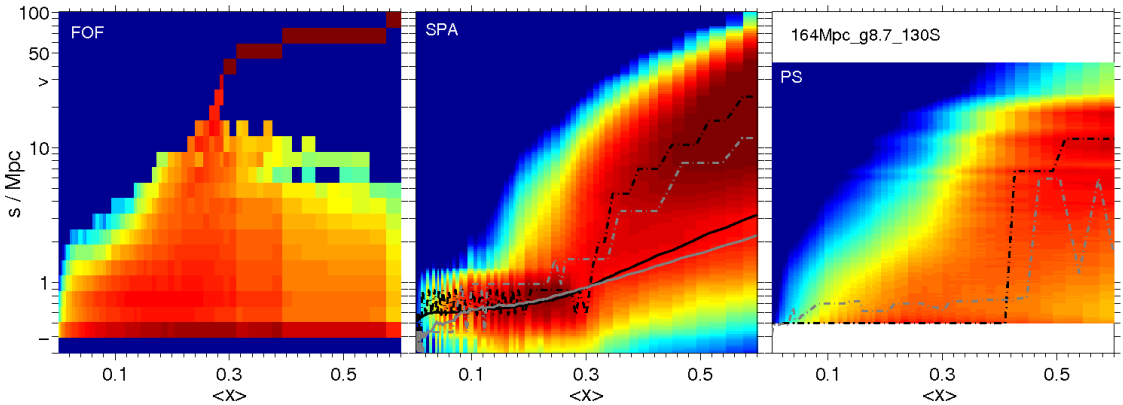}
\caption{Color plots of evolution of size distributions for simulation 
163Mpc\_g8.7\_130S, from left to right: FOF, SPA (including $3V/A$, solid 
black line) and PS. Color coding and line-styles have the same meaning as 
in Fig. \ref{fof_comp2}. 
\label{scale_comp2}
}
\end{center}
\end{figure*}

Equipped with an idea about which effects can be due to the changed resolution,
 we now turn to the  larger volume simulation. 
In Fig. \ref{scale_comp2} we plot all 4 size measures for simulation 
163Mpc\_g8.7\_130S. We concentrate first on the $3V/A$ size estimate, see the 
middle panel of that figure (black solid line) and compare it to the $3V/A$ 
estimate of the fiducial simulation (gray solid line): except for very low 
global ionization fractions ($\left<x\right> \leq 0.08$) where the estimate 
for the fiducial simulation is of the order of the cell size of the 163~Mpc
 simulation volume, the two curves almost coincide up to a global ionization 
fraction $\left<x\right> \sim 0.3$, after which the scale in the 163~Mpc 
simulation grows faster. 

The SPA distribution shows a similar behaviour as $3V/A$: below ionization 
fractions $\left<x\right> \sim 0.08$, the SPA peaks of the fiducial 
simulation are at scales comparable to the cell size of the 163~Mpc simulation 
while the peaks of 163~Mpc simulation are slightly larger (compare the 
dot-dashed lines in the same panel). The fiducial simulation shows also a 
wider distribution with contributions from smaller as well as from larger 
scales, see left panel in Fig.\ref{spa_comp}. Between 
$\left<x\right> \sim 0.08 \-- 0.3 $ the evolution of SPA distribution of the 
two simulation is very similar. At larger global ionization fractions, the 
scale of the SPA peak is growing faster in the 163~Mpc simulation. 

The power spectrum also shows that, below global ionization fractions 
$\left<z\right> \sim 0.3$, despite the different resolution, the simulations
 with different simulation volumes agree well. At global ionization fractions
 $\left<x\right> \ge 0.4 $ there is considerable power on scales that are not
 captured in the 53~Mpc simulation which might explain the shift found by the
 other size distribution estimates. The power spectrum suggests that the
 163~Mpc simulation captures the most relevant scales involved in reionization
 (that is H~II region sizes up to roughly $\left<x\right> \sim 0.5$, 
  H~I region sizes above this)
  since there is little power on scales above  $s \sim 30$~Mpc.  
 A noticeable difference between the simulations is the lack of the peak at
 scales around 0.8~Mpc in the large simulation volume. As found earlier by
 the smoothing test, this may be a resolution effect. The peak at scales 
$s \sim 6$~Mpc is not as clear in the 163~Mpc volume as in the 53~Mpc 
volume. After $\left<x\right>\sim 0.5$ it shifts to larger scales. 

The FOF size distribution (left panel of Fig. \ref{scale_comp2} and left 
panel of Fig.\ref{fof_comp}), of the two simulations look on first sight 
very different. It can be easily understood why: the smallest scale of H~II
 regions in the small simulation volume is smaller than that in the large 
simulation volume. Looking at the cell-volume limits which are indicated 
on the abscissa, it can be seen that the additional population of small
 scale H~II regions present in the 53~Mpc simulation is below the cell-size 
of the 163~Mpc simulation and therefore below its resolution limit. All 
those H~II regions are partly ionized in the large simulation volume. Therefore,
 some of them which are more ionized than $x_{\rm th}=0.5$ appear as additional 
population in the FOF distribution of the large simulation volume at 
scales of its cell size. At 10\% global ionization fraction, it can be seen 
that the 53~Mpc simulation has slightly larger large-scales than the 163~Mpc
 simulation, $s \sim 4 $Mpc and $s \sim 5$Mpc, respectively. Partly ionized 
cells that are ionized below the threshold value in the larger simulation 
volume and therefore do not count as belonging to the H~II region cannot 
account completely for this difference in size. This slight mismatch in size 
between large and small simulation volume might be an artifact from our 
implementation of suppression: since we suppress all sources inside a cell 
that has a mass averaged ionization fraction $\bar x_m$ larger than 10\%, 
the volume in which sources are suppressed can be overestimated in 
simulations with larger cell sizes.

While the size distribution in the small simulation volume shows a gap 
after $\left<x\right>\sim 0.1$, 
the gap emerges first at $\left<x\right>\sim 0.25$ in the large simulation 
volume. This is because a single region in 
any of the bins that are empty in the small simulation volume but populated 
in the large one, would already 
exceed the contribution it makes in the small volume. This sampling effect 
was already mentioned in the previous section. 

\begin{figure}
\begin{center}
  \includegraphics[width=4cm]{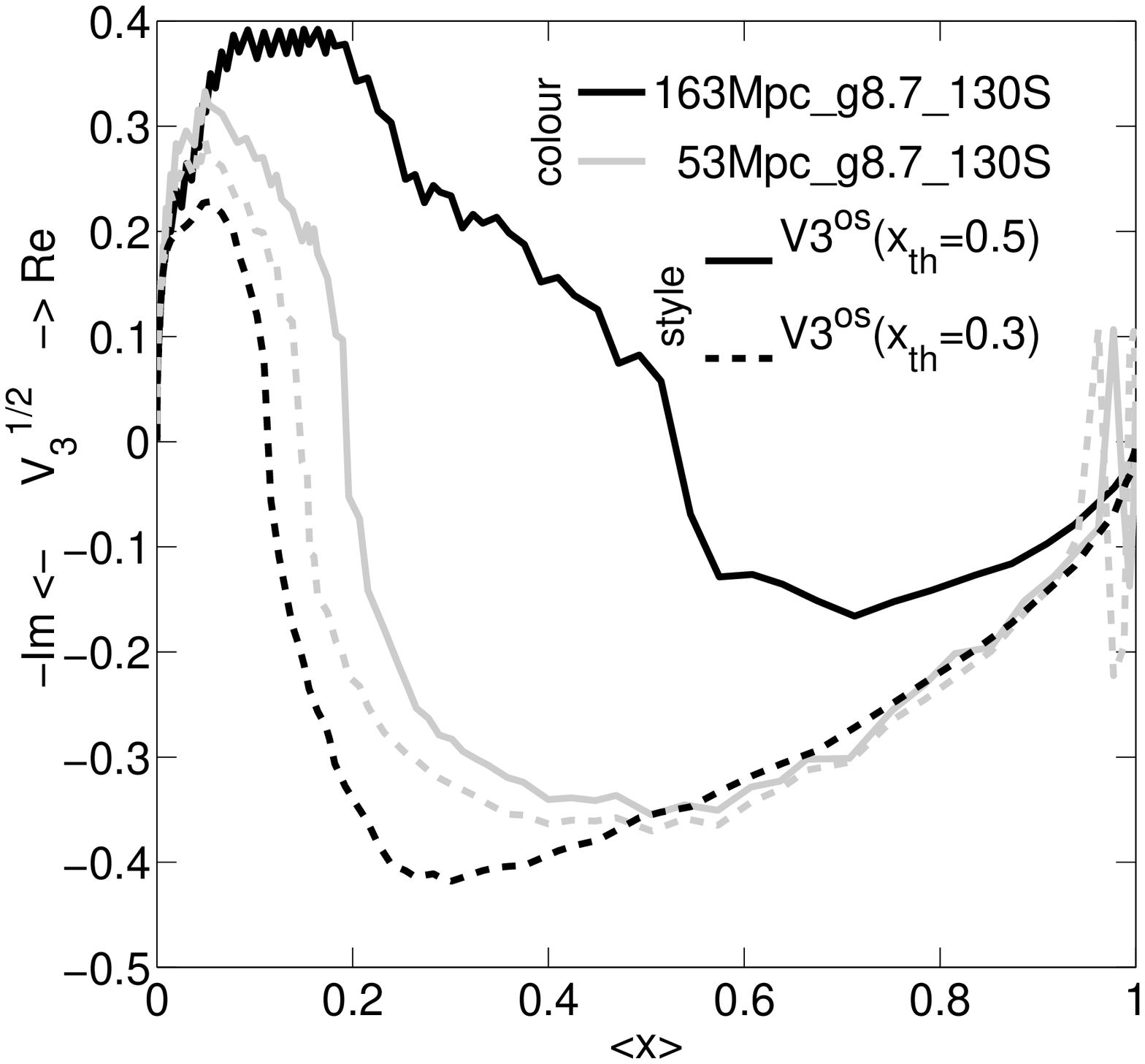}
  \includegraphics[width=4cm]{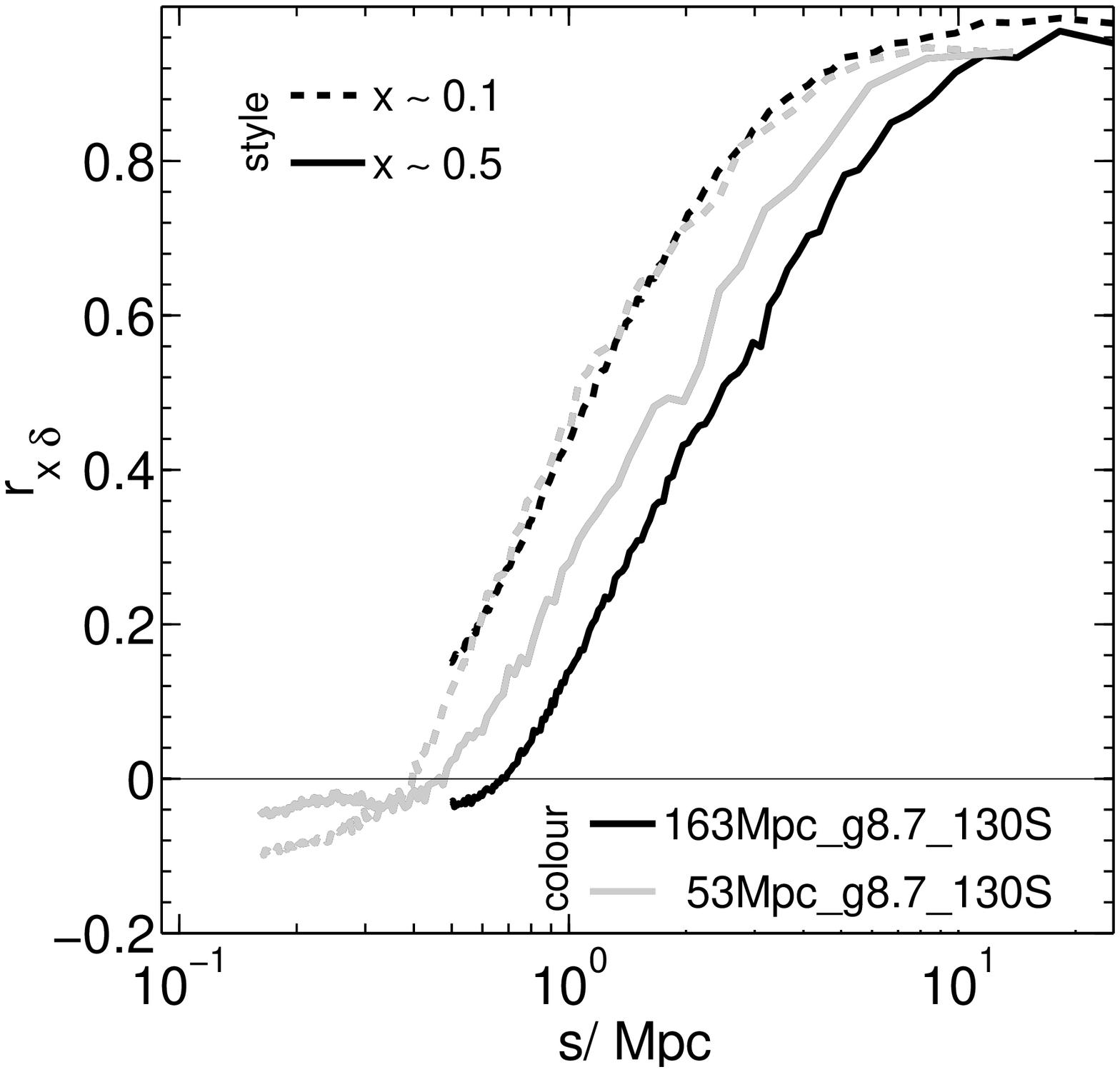}
\caption{\textbf{Left panel:} Evolution of Euler characteristic $V_3$ at 
threshold value $x_{\rm th}=0.5 $ and 0.3 as indicated in the figure (by line style) 
for both g8.7\_130S simulations, gray lines indicating the 53~Mpc simulation 
and black lines the 163~Mpc simulation. \newline
\textbf{Right panel:} Cross correlation of the density field with ionization 
fraction of both g8.7\_130S simulations (with the same color coding as in the
 left panel) at two different global ionization fractions as indicated in
 the figure (by line style). 
\label{mink2} 
}
\end{center}
\end{figure}

At 30 \% global ionization fraction, the ionized volume of the single large 
connected region in the 163~Mpc simulation is approximately 10\% of the total 
simulation volume which is larger than the size of the 53~Mpc volume. Similarly, 
the volume of the largest connected ionized region in the 53~Mpc simulation is 
also about 10\% of the total volume. 
The fact that the largest connected region is a constant fraction of the simulation 
volume already at 30\% ionized fraction suggests that this  region pervades the 
whole simulation volume. 
This statement is strengthened by the Euler Characteristic of these simulations: 
$V_3$ is already highly negative at $\left< x \right> \sim 0.3$ (for the 163~Mpc  
simulation this is only true for lower threshold values), see left panel of Fig. 
\ref{mink2}. It should be noted that $V_3$ of the 163~Mpc simulation shows 
very similar evolution to $V_3$ of the 53~Mpc simulation (for a low threshold 
value for the 163~Mpc volume). The fact that $V_3$ in the large simulation 
volume is highly dependent on threshold value, shows that the resolution 
is not sufficient to use $V_3$ as a reliable analysis tool: the ambiguity as to
 whether regions are connected or not is too high, as can be seen comparing the 
curves for different threshold values in the left panel of Fig.\ref{mink2}. 
However, it can be seen that the effect of including "lost" photons, is much 
smaller in the large simulation volume: relatively fewer photons travel distances 
longer than a box distance in the large simulation volume. 

The cross correlation between ionized fraction and density at low global ionization
 fractions is almost identical for both simulation volumes as the thin lines in 
the right panel of Fig. \ref{mink2} show.  It differs more at higher global 
ionization fractions where it shifts towards larger scales for the 163~Mpc 
simulation, probably indicating that the scales dominating the ionization 
field at that global ionization fraction exceed the size of the  53~Mpc simulation.

%-------------------------------------------------------------------------------
%-------------------------------------------------------------------------------
%-------------------------------------------------------------------------------
% SECTION 5--------------------------------------------------------------------
\section{Physical parameters}  
Our efficiency parameter $g_{\gamma}$  is a product of the efficiency of star 
formation, production of ionizing photons per stellar atom (related to the initial 
stellar mass function) and the escape fraction of the photons from the galactic 
halo into the intergalactic medium. All these quantities are not very well 
constrained  at present. Also the efficiency of suppression due to 
Jeans-mass filtering can be different from the simple on-off function as implemented 
in our simulations with suppression
\citep[c.f][]{2007MNRAS.377.1043M,2008MNRAS.390.1071M,2008MNRAS.390..920O}. To 
study the effect of our simplified suppression model, we consider in this section 
some extreme scenarios and use the methods described above to investigate the 
effect on the scales and the topology of the 
emerging H~II regions.  
\subsection{Minimum mass of halos hosting sources with escaping ionizing radiation}
\begin{figure*}
\begin{center}
  \includegraphics[width=15cm]{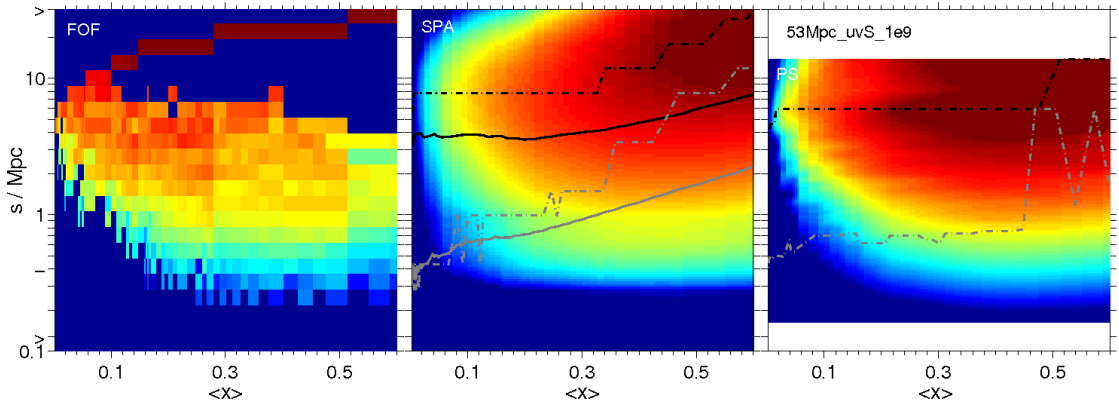}
\caption{Color plots of evolution of size distributions for simulation 
53Mpc\_uvS\_1e9, from left to right: FOF, SPA (including $3V/A$, solid black line) 
and PS. Color coding and line-styles have the same meaning as in Fig. \ref{fof_comp2}. 
\label{fof_comp3}
}
\end{center}
\end{figure*}
\begin{figure*}
\begin{center}
  \includegraphics[width=15cm]{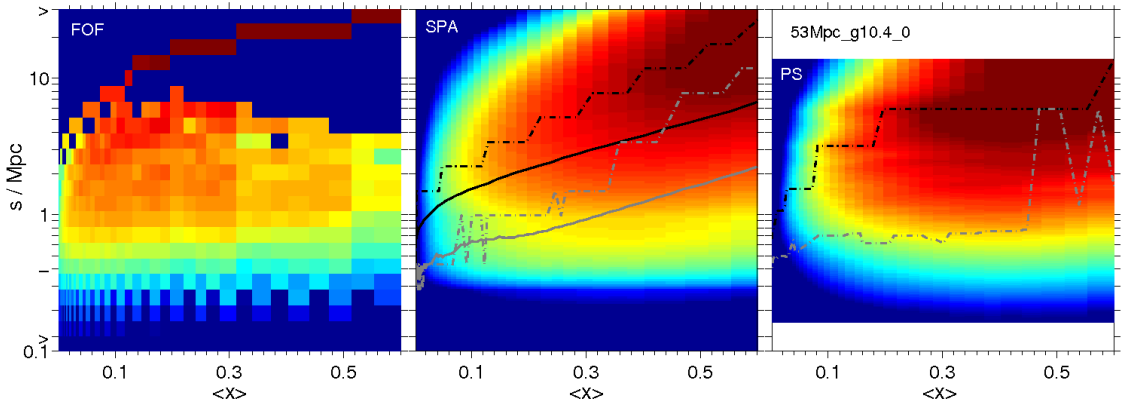}
\caption{Color plots of evolution of size distributions for simulation 
53Mpc\_g10.4\_0, from left to right: FOF, SPA (including $3V/A$, solid black line) 
and PS. Color coding and line-styles have the same meaning as in Fig. \ref{fof_comp2}. 
\label{fof_comp4}
}
\end{center}
\end{figure*}
In this subsection we investigate how a change in the source population 
affects the simulation. In simulations 53Mpc\_uvS\_1e9  and 
53Mpc\_g10.4\_0 only halos more massive than $10^9 M_{\odot}$ host sources 
that  emit ionizing radiation into the IGM.
The former simulation is constructed such that the number of released ionizing photons 
in every time step (after the formation of the first massive halos) is the same 
as for our fiducial simulation. The sum of all photons that were released in the 
fiducial simulation by low mass halos up to the time at which the first massive 
halo is forming, is emitted additionally in the first time step after which the 
first massive source has formed. Since the number of forming halos increases 
exponentially, the fraction of additionally released photons in this first time 
step is only about half of the total released photons at that time step. As 
pointed out in section \ref{sims}, the resulting source efficiency is variable 
with time, shown in Fig.\ref{ggamma_evol}. This also means that the minimum number 
of photons released by one source during one time step is decreasing with 
increasing $\left< x \right>$. Therefore, also the minimum size for H~II regions 
decreases   to $\left<x\right> \sim 0.25$. This can be seen most clearly in the 
FOF size distribution (Fig.~\ref{fof_comp3}, left) and in the power spectrum 
(Fig.~\ref{fof_comp3}, right). To avoid this effect we performed simulation 
53Mpc\_g10.4\_0 which has a different ionization history from our fiducial 
simulation, but the source efficiency of the high mass sources is chosen such 
that overlap occurs at roughly the same time, as can be seen in table 1. 

The FOF size distribution shows that individual H~II regions grow larger before
 merging with the largest H~II region in both, the 53Mpc\_uvS\_1e9 and 
53Mpc\_f10.4\_0 simulations than in the fiducial one; the gap in the distribution 
is smaller. This is due to the greater average distance between high mass sources. 
The space in between the large H~II regions is neutral, without any ionized spots. 
Therefore, each individual H~II region can grow bigger before merging. 

 Below global ionization fractions $\left< x \right> \sim 0.2$, the evolution of
 sizes in the 53Mpc\_uvS\_1e9 simulation is dominated by the first H~II regions 
emerging around the highly efficient first sources. At higher global ionization 
fractions the size-evolution is very similar to simulation 53Mpc\_g10.4\_0. 
Therefore, we concentrate in the following on the 53Mpc\_g10.4\_0 simulation.

Compared to our fiducial simulation, the 3V/A estimate of simulation 
53Mpc\_g10.4\_0  (middle panel of Fig.\ref{fof_comp4}) suggests an average bubble 
scale about a factor 3 greater at all global ionization fractions we consider here. 
Also the spherical average distribution clearly shows this shift to larger scales. 
This is best visible when comparing the peak-scales (dot-dashed curves in the 
same panel). The power spectrum (Fig. \ref{fof_comp4}, right panel), reveals 
that it is only a shift to bigger scales below global ionization fractions of 
about 20\%. The H~II regions that form first seem to be larger than the ones 
in the fiducial simulation. Later, it is rather a lack of small scales; notably 
the peak at scales $s \sim 0.8$~Mpc is absent. This is not surprising as we 
identified the peak to be due to the suppression of sources in low mass halos. The suppresion
is responsible for halting the growth of the H~II regions formed by these sources completely.
The peak at scales $s \sim 6$~Mpc is still there. However, there is more power 
on scales that are not captured by the 53~Mpc simulation volume. 

The Euler Characteristic for simulation 53Mpc\_g10.4\_0, see light gray lines 
in the left panel of Fig. \ref{dxx_comp3},  is in total much flatter than the 
Euler Characteristic for the fiducial simulation. Since we saw in the FOF 
distribution and the power spectrum that there is not much contribution from 
scales that are hardly resolved, this is very unlikely to be due to unresolved 
scales. We therefore conclude that the bigger H~II regions around rare sources 
result in a less complex topology with fewer tunnels and cavities. 
We saw above that feeding back diffuse photons into the volume, affects $V_3$ 
at high global ionization fractions. The fraction of those photons for 
simulation 53Mpc\_g10.4\_0 is already 0.2 at 80\% global ionization fraction. 
Therefore, the evolution of $V_3$ beyond $\left<x\right> \sim 0.8$ might be
 dominated by the effect of these photons, as described in section \ref{ana_method_intro}. 
It can be seen that $V_3$ for the low threshold value is at some points 
considerably different from the value at $x_{\rm th}=0.5$. However, since we plot
 the square root of $V_3$, the differences at values close to 0 are amplified. 
The actual difference is never greater than 20\% of the maximum value.  
\subsection{Source suppression vs. low efficiency}
To test the effect of source suppression in regions where the IGM is ionized, 
we compare our fiducial simulation (with instantaneous complete suppression of 
sources in low mass halos in ionized regions) to two simulations without suppression: 
53Mpc\_g8.7\_130 which has the same source efficiencies but which ends much 
earlier due to the many more released photons and 53Mpc\_g0.4\_5.3 which has 
substantially lower source efficiencies to end at roughly the same time as 
the fiducial simulation. 

\begin{figure*}
\begin{center}
  \includegraphics[width=15cm]{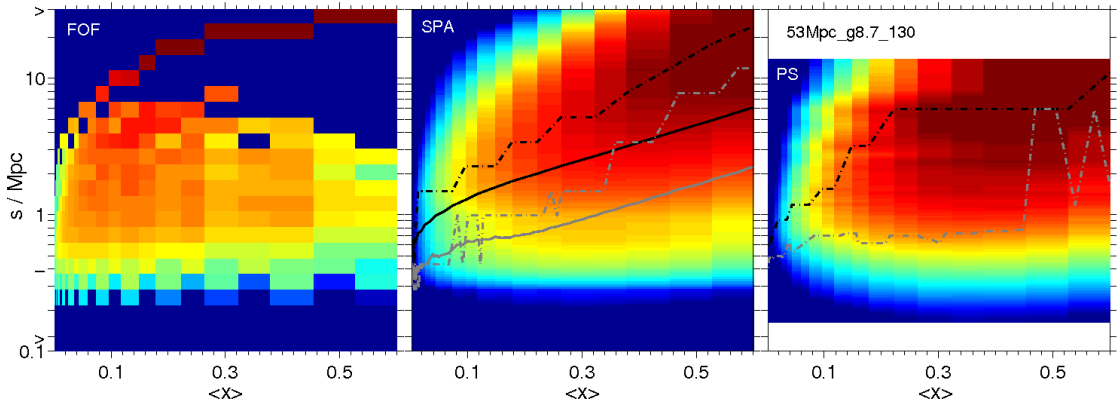}
\caption{Color plots of evolution of size distributions for simulation 
53Mpc\_g8.7\_130, from left to right: FOF, SPA (including $3V/A$, solid black line) 
and PS. Color coding and line-styles have the same meaning as in Fig. \ref{fof_comp2}. 
\label{fof_comp4a}
}
\end{center}
\end{figure*}

\begin{figure*}
\begin{center}
  \includegraphics[width=15cm]{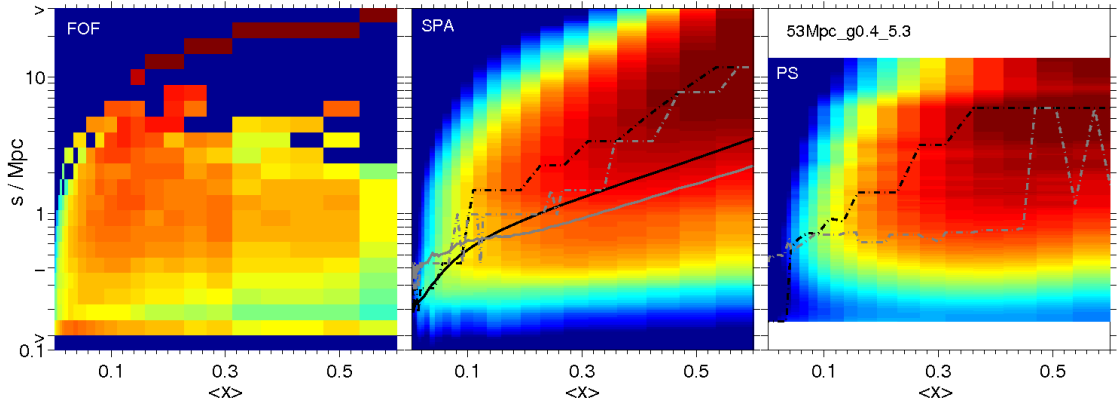}
\caption{Color plots of evolution of size distributions for simulation 
53Mpc\_g0.4\_5.3, from left to right: FOF, SPA (including $3V/A$, solid black 
line) and PS. Color coding and line-styles have the same meaning as in Fig. \ref{fof_comp2}. 
\label{fof_comp4b}
}
\end{center}
\end{figure*}

The comparison of the FOF size distributions between model 53Mpc\_g8.7\_130 
and 53Mpc\_g0.4\_5.3 (see left panels in Fig. \ref{fof_comp4a} and Fig. 
\ref{fof_comp4b}) shows that the model with the lower source efficiencies 
shows more very small H~II regions than the fiducial simulation, similar to 
53Mpc\_g1.7\_8.7S. The simulation with the same source efficiencies as the 
fiducial simulation but without suppression shows less small H~II regions 
than the fiducial simulation, because each individual source forming is active
 longer and so continuously grows its H~II region. Also, clustered sources in
 the same or neighboring cells support the growth of their joined H~II region. 

The spherical average distribution, see middle panels of Fig. \ref{fof_comp4a}
 and Fig. \ref{fof_comp4b} for the higher and lower efficiency simulations 
without suppression, respectively, show a very similar evolution. The rate
 at which the average size grows seems to be the same, but the scale is
 shifted towards larger scales for the simulation 53Mpc\_g8.7\_130 by about a 
factor 1.5. Also, the $3V/A$ size estimates suggests an almost constant shift to larger scales. 

The power spectra, see right panels of the same figures show more power on 
small scales (below $s\sim 0.5$) and less power on large scales 
(above $s\sim 6$) for 53Mpc\_g0.4\_5.3 than for simulation 53Mpc\_g8.7\_130. 
However, up to global ionization fraction $\left<x\right> \sim 0.5$, the peak 
at scales around $s \sim 6$~Mpc is present in both simulations. The peak at
 scales $s \sim 0.8$~Mpc is absent in both simulations.

It should be noted that the size distributions found by all three methods 
as well as the 3V/A size estimator of the 53Mpc\_g8.7\_130 simulation are 
very similar to the ones from 53Mpc\_g10.4\_0. However, there is a very 
significant shift in time between these two simulations. The reason for their
 similarity if compared at equal $\left<x\right>$ might be that most halos 
which first reach masses above $10^8 ~M_{\odot}$, reach masses above  $10^9 ~M_{\odot}$ 
at accordingly lower redshifts. Therefore, the same halos hosting sources 
which turn on as sources in low mass halos in the former simulation, turn on
 as sources in more massive halos at a later time. This means that if 
suppression of sources is not important, the sources ``shaping'' reionization are 
the ones in halos with the lowest mass that can form luminous sources. 

\begin{figure}
\begin{center}
  \includegraphics[width=3.9cm]{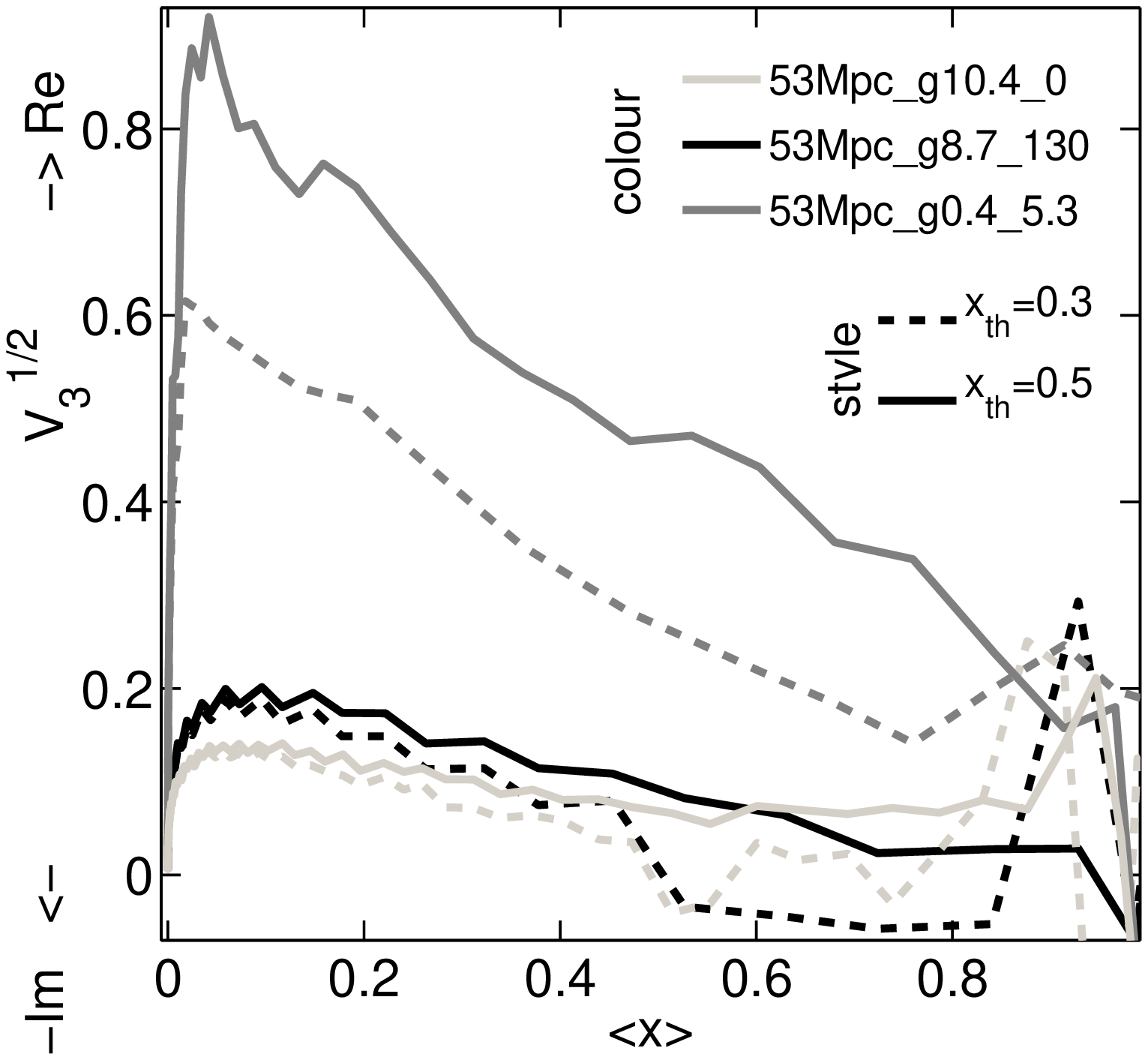}
  \includegraphics[width=3.9cm]{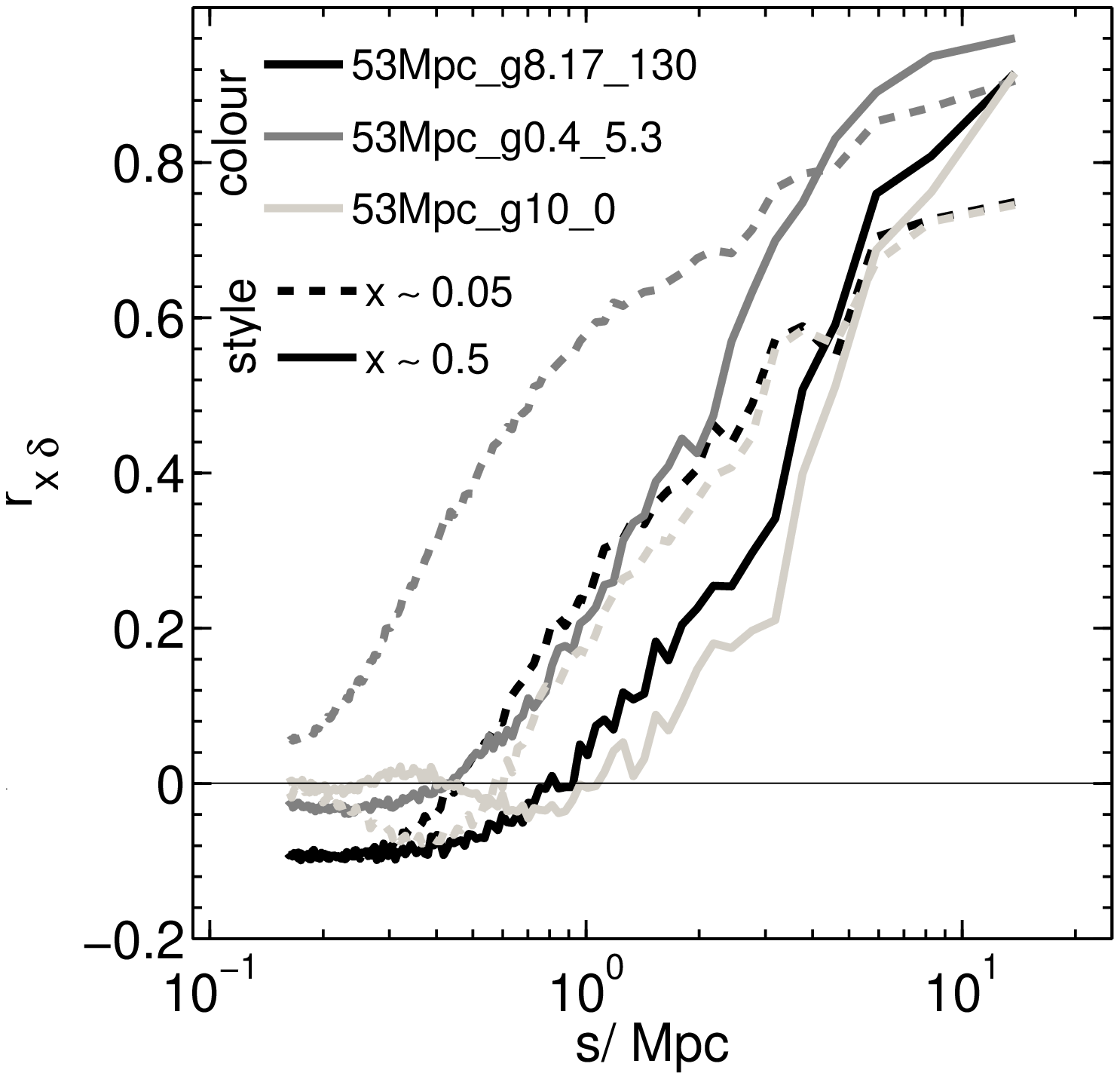}
\caption{\textbf{Left panel:}$V_3$ of simulation 53Mpc\_g10.4\_0, 53Mpc\_g8.7\_130
  and 53Mpc\_g0.4\_5.3, as indicated in the figure (by line color)  at threshold 
values $x_{\rm th}= 0.5$ and 0.3, as indicated in the figure (by line style). \newline
\textbf{Right panel:} Cross-correlation coefficient for the same simulations 
at different $\left< x \right>$, as indicated in the figure (by line style). 
\label{dxx_comp3}
}
\end{center}
\end{figure}

The Euler Characteristic (left panel in Fig. \ref{dxx_comp3}) shows that $V_3$ 
never becomes negative for both, simulation 53Mpc\_g0.4\_5.3 (dark gray lines) 
and  simulation 53Mpc\_g8.7\_130 (black lines): There are many more disconnected
 H~II regions than neutral tunnels through the ionized regions. For a lower threshold 
value, $x_{\rm th}=0.3$, $V_3$ goes mildly negative. The high dependence of 
simulation 53Mpc\_g8.7\_130 on threshold value is somewhat surprising since 
the smallest H~II regions are larger than for the fiducial simulation. 
However, the total number of H~II and H~I regions is small (only a few hundred
 compared to a few thousand for the fiducial simulation), therefore a few 
critical connections are enough to introduce a strong dependence of $V_3$ on threshold value.
For the 53Mpc\_g0.4\_5.3 simulation it can further be seen that the maximum
 $V_3$ is about a factor of four larger (at low threshold value which means 
most probably overestimating the connections of regions) than for the fiducial
 simulation. There are more disconnected H~II regions due to their smaller 
minimum sizes. Simulation 53Mpc\_g8.7\_130 reaches the same global ionization 
fraction as simulation 53Mpc\_g0.4\_5.3 at a much higher redshift, at a time 
where fewer halos have formed. The maximum $V_3$ is smaller for the
 53Mpc\_g8.7\_130 simulation because the individual H~II regions grow 
bigger and merge earlier in terms of global ionization fraction. However, 
the big differences between $V_3$ at different threshold values show that
 these simulations do not have sufficient resolution, therefore the ambiguity 
of connections is too high. 

Fig. \ref{dxx_comp3} (right panel) shows that the correlation between ionized 
fraction and overdensity at small global ionization fractions, here represented 
by $\left<x\right> \sim 0.05$, is flatter for  simulations without suppression, 
or simulations without any suppressible sources, like simulation 53Mpc\_g10.4\_0:
 on large scales, the correlation is lower than for simulations with suppression 
like our fiducial simulation; on small scales, relatively larger. This behaviour 
can be interpreted together with the results from the Euler Characteristic in the 
following way: The lack of suppression leads to earlier break outs of the ionizing 
radiation into the low density regions between low mass sources  which reduces 
the number of neutral tunnels and  lowers the cross-correlation on larger scales.
At the same time, not suppressing sources in partly ionized regions leads to 
complete ionization of that region which increases the cross-correlation on smaller scales. 

At higher global ionization fractions, here represented by $\left<x\right> \sim 0.5$,
 these differences disappear because almost all sources in low mass halos are 
suppressed in the simulations with suppression and the reionization process is 
dominated by the sources in high mass halos. 

\cite{2007MNRAS.377.1043M} tested the effect of suppression for the case of 
equally efficient high and low- mass sources. However, their simulations do not 
resolve halos below $10^9 \rm{M}_{\odot}$. They use an analytic prescription to 
include unresolved halos above the H~I atomic cooling mass. Among other things 
they found that even their most drastic (instantaneous) suppression model 
(complete suppression of sources in low mass halos) yields an ionization 
field with similar morphology to a simulation without suppression (their
 simulations F3 and S1, respectively). Additionally they tested a simulation
 with higher efficiencies for low mass sources but without suppression
 (their model S2). To realize this they chose a mass dependent source efficiency
 factor. Very roughly, this translates into $(10^8/10^9)^{-2/3}\approx 5$ times 
more efficient low ($10^9 > M/\rm{M}_{\odot} \ge 10^8$) than high 
($M/\rm{M}_{\odot} \ge 10^9$) mass sources in terms of our step-function
 efficiency assignment method (our fiducial model has 15 times more efficient 
 sources in low mass halos). They found that the effect of boosting the 
efficiency of low mass sources on ionization field morphology is rather small. 
From these two tests (F-set vs. S1 and S1 vs. S2) they concluded that suppression,
 even in the case of high efficient low mass sources, does not affect the 
morphology of the ionization field. We can confirm that the ionization field morphology 
does not change much if the efficiency of low mass sources is boosted, as long as the 
total photon output of the least luminous sources that form the smallest HII 
regions is held roughly constant (compare for example 53Mpc\_g10.4\_0 and 53Mpc\_g8.7\_130, 
Fig. \ref{fof_comp4} and Fig. \ref{fof_comp4a}, respecively). 
However, comparing our fiducial simulation (with suppression) 
to simulations 53Mpc\_g8.7\_130 and 53Mpc\_g0.4\_5.3 (both without suppression), 
we find large differences in the topology and in the average size distributions 
of ionized regions, as discussed above. This is most probably due to the 
early break out of H~II regions formed around sources in low-mass halos in the case 
of no suppression and high low-mass halo source efficiency.

%-------------------------------------------------------------------------------
%-------------------------------------------------------------------------------
%-------------------------------------------------------------------------------

% DISCUSSION
\section{Conclusions}  

We have explored different methods for characterizing the scales and
topology of complex ionization fraction fields produced by simulations
of cosmic reionization. For characterizing the length scales or sizes
of HII regions, we used three methods that give a distribution of
scales; the FOF method, the spherical average method and the power
spectrum of the ionization fraction field. In addition we proposed a
single valued measure of average size of HII region, given by the
ratio of the volume to the surface of all regions. For characterizing
the topology we employed the Euler Characteristic or third
Minkowski functional, $V_3$ of the ionization fraction field.

The nature of the size distribution of H~II regions can be viewed to
be a matter of definition. Applying a literal definition leads to the
FOF approach, in which the HII regions are considered to be connected
regions of space.  Because the topology of reionization can be quite
complex, as seen from the Euler characteristic, this definition does
not lend itself easily to analytical modeling, and connecting the FOF
size distribution to other scale estimators, such as the power
spectrum, is by no means trivial. For the FOF method, what is lost in
its complexity is gained in the detailed description of the
reionization process that it provides. Although most of the volume is
already at quite low global ionization fractions contained in one
large connected region, there is a wealth of information contained in
the number and sizes of the smaller bubbles, which only occupy a small
fraction of the volume. As we found in section \ref{numericals}, the
FOF size distribution is affected by resolution in terms of cell size and one has to
take this into account when interpreting the results and comparing
simulations with different resolution. In principle, the FOF method
yields the maximum size for isolated ionized regions, before they
start to merge into the largest one. However, in practice, this scale
is dependent on the sampling of the distribution function and
therefore on the size of the simulation volume.

The spherical average method, on the other hand, gives distributions
which are much smoother and can more easily be connected to analytical
models \citep{2007ApJ...654...12Z}. The results are more sensitive to
the large scale H~II regions which constitute the main contribution to
the global ionization fraction. Due to its averaging nature, it washes
out the details. Consider, for example, a collection of two types of
spheres with scales $s_1=a$ and $s_2=b>a$; the spherical average
method will only reveal the two distinct scales if $a/b \le
1/3$. Also, the spherical average alone tends to substantially
underestimate the sizes of H~II regions, as shown in
Fig. \ref{spa_test} by our toy model.  For using it as a size
estimator on par with other methods, the spherical average scales
should be multiplied by a factor 4.

When the universe is mostly neutral, the peak of the ionization power
spectrum is related to the size distribution of {\em ionized}
regions. For a single top-hat sphere, the first peak in the power
spectrum is related to its radius by $k_{\rm max}= 2.46/r$ and we
choose to use $2.46/k$ as a size estimator when comparing to the
results of other methods. The power spectrum produces size
distributions roughly comparable to those from the spherical average
method. However, the advantage of the power spectrum over the
spherical average method is that it does not wash out the details of
the distribution. As mentioned earlier, of all size estimators discussed in this paper,
the power spectrum is the one most related to upcoming observations as
one component in the expansion of the 21 cm power spectrum \citep[e.g.][]{2006PhR...433..181F}.

The ratio of the zeroth and first Minkowski functionals can be used to define
a mean radius of HII regions, the $3V/A$ estimate. This estimate is a 
surface weighted average. It generally gives results consistent with
the maximum of the spherical average, and when it falls below that
there is a large fraction of small bubbles (which dominate the surface).

The Euler characteristic $V_3$ of the ionization fraction field offers
a rich description of the evolution of the topology of
reionization. Taking our fiducial simulation as an example, in the
early stages of reionization the value of $V_3$ is positive as the
topology is dominated by a large number of isolated H~II regions.
However, already beyond global ionization fractions of roughly 20\%,
$V_3$ becomes highly negative, indicating a complex topology of
connected HII regions with tunnels. This is consistent with the
distribution from the FOF method, which shows that already at $\left<
  x \right> \sim 0.3$ the main contribution to ionized fraction comes
from only one large connected region which pervades most of the
simulation volume. 

As the ionization fronts around stellar sources are quite thin,
ideally $V_3$ should not depend much on the chosen threshold value, and
the most interesting aspect is the change of $V3$ with time. In this
sense topology studies of HII regions differ from those of density
fields, where the variation of $V_3$ with threshold value is used to
characterize the field at a given time. However, in practice the
resolution of the ionization field may not be sufficient to achieve
sharp fronts, and partially ionized cells will occur. The result is
different values of $V_3$ for different threshold values and even
different evolutions of $V_3$ at different threshold values. More
seriously is the interaction of this effect with the definition of
connectivity/adjacency used when calculating the Euler Characteristic.
We proposed a new test for establishing how robust the derived values
of $V_3$ are to a change of adjacency. In this one compares the answer
with the value obtained for a field and threshold value which
have their sign inverted. Using this we showed that sub-sampling or
smoothing is generally required to obtain consistent results, and that
for fields with a large fraction of partially ionized cells, it can
be difficult to get consistent results.

We subsequently applied the size estimators and Euler Characteristic to
study differences and similarities between different reionization
simulations. Comparing identical simulations in two different volume
sizes, 163 Mpc and 53 Mpc, shows that below global ionization
fractions of 30\% the average scales of H~II regions are roughly the
same for both simulations. Beyond that the size distributions in the
larger volume start to contain scales beyond those available in the
smaller one. Another manifastation of this is that the largest connected region 
found by the FOF-method for both simulation volume sizes is 10\% of the each of 
the simulation volumes already at 30\% global ionization fraction. 
Therefore, this largest H~II region is about a factor 
30 larger in the large simulation volume: A region of this size does not fit 
into the 53~Mpc simulation volume. 

Even earlier there are differences between the
two FOF distributions showing an absence of H~II regions with volumes
of a few hundred Mpc$^3$ in the small box. This is because relatively
isolated overdensity regions (surrounded by larger voids) are missing
in the small box due to numerical variance. Since the contribution
from those scales is very small, the effect on the average scales is
negligible. In general, simple size estimates may be biased due to
missing scales (e.g. intermediate scales missing in the small box due
to the smaller sampling volume, small scales missing in big box due to
resolution). If one is not interested in very small scales, the lower
resolution in the big box appears not to be a problem.  The fact
that the position of the peak of the power-spectrum at all times
(global ionization fractions) is well below the box-size scale for the
163~Mpc box, indicates that no larger simulation boxes are needed to
follow the evolution of the peak position.

Comparing simulations with and without suppression (either with only
high mass sources, or with sources in low mass halos not suppressed in
ionized regions) shows that the ones without suppression typically
have a much less complex ionization fraction field topology and a more steady growth
of average H~II region sizes. We thus find significant differences
between the cases with and without suppression. For the simulations
with sources only in high mass halos, this can be explained as
follows: the individual H~II regions can grow much bigger before
merging, due to the larger average distances between high-mass
sources. For the simulation without suppression and high efficiency
sources a similar argument holds, but the whole process takes
place at much earlier redshifts. For the simulation without
suppression and with low efficiency sources the explanation is
again similar, but now applied to smaller scales instead of at earlier
times. This similarity between the cases without suppression is due to
the statistical nature of the density field. If suppression is not
important, the sources in halos with the lowest mass still capable of
forming sources are the ones shaping reionization.

By imposing an external photon-budget on an independently evolved
source population, one can in principle separate the effect of the
source population from that of the reionization history. However, by
necessity this implies an evolution of the source efficiencies, which
in the case we studied seriously affected the evolution sizes of HII
regions. Up to a global ionization fraction $\left<x \right> \sim 0.2$
the size distribution is dominated by the size of H~II regions
generated by the first generation of sources. In the later stages of
reionization, the morphology of the ionization field is very similar
to a simulation with a fixed source efficiency for high mass sources.

As outlined in the introduction, we concentrated on the analysis of
the ionization fraction fields and its evolution. Applying the various
analysis methods to the future observations of the redshifted 21cm
signal is in principle possible, but requires sufficient sensitivity
to image the signal at different frequencies. The first generation of
telescopes is not expected to be able to do this, but the planned
Square Kilometer Array (SKA) \footnote{http://www.skatelescope.org}
should be. Compared to the simulation results, the observations will
have the additional complications of noise and limited spatial
resolution (below even that of our 163~Mpc simulation). As we have
shown, resolution effects should be treated with care, especially for
the Euler Characteristic, and noise peaks can obviously also bias the
topology determination. Still, characterizing the morphology of HII
regions in the data will be important as they trace the mass and thus
the emerging cosmic web. We leave the application of the various
size/scale and topology estimates to mock observational data to a
future paper.
%-------------------------------------------------------------------------------
%-------------------------------------------------------------------------------

%-------------------------------------------------------------------------------
% ACKNOWLEDGEMENTS--------------------------------------------------------------
\section*{Acknowlegments} 
We are grateful to Thomas Buchert for providing us with the code used
to compute Minkowski functionals. A significant fraction of the RT simulations 
were run on Swedish National Infrastructure for Computing (SNIC) resources at HPC2N 
(Ume\aa, Sweden). The authors acknowledge the Texas Advanced Computing Center
(TACC) at The University of Texas at Austin for providing
HPC resources that have contributed to the research results
reported within this paper. URL: http://www.tacc.utexas.edu
This work was supported in part by Swedish Research Council grant 2009-4088, NSF 
grant AST 0708176, NASA grant NNX07AH09G, Chandra grant SAO TM8-9009X, and 
NSF TeraGrid grants TG-AST0900005 and TG-080028N.
%-------------------------------------------------------------------------------
%-------------------------------------------------------------------------------
%\bibliography{refs}
%\bibliographystyle{mn2e}

% APPENDIX----------------------------------------------------------------------
% SPA---------------------------------------------------------------------------
\begin{appendix}
\section{Spherical average size distribution of a log normal distribution}
\label{app_spa}
The spherical average method was described by \cite{2007ApJ...654...12Z}. 
In this technique, each cell in the computational volume is considered to be in an
ionized region if a sphere centered on that cell has a mean ionized fraction
greater than a given threshold, usually $x_{\rm th}=0.9$. The size of the H~II
region to which it belongs is taken to be the largest such sphere for which
the condition is met.

In the following, we assume that gas is either fully
ionized or fully neutral, and that all ionized bubbles
are non-overlapping spheres with a volume-weighted distribution $dP/dR$,
so that $P(R+dR)-P(R)$ is the fraction of the ionized volume that lies
within bubbles with radii between $R$ and $R+dR$. 
What bubble distribution, $dP_{\rm sm}/dR$, would be obtained by using the
spherical average method? To simplify further, we will
take the threshold for the spherical average, $x_{\rm th}$, to be
arbitrarily close to unity, so that a point is considered to be within
an ionized sphere of a given radius only if {\em all} the matter in
that sphere is ionized. For a single ionized sphere of radius $r$, 
$dP_{\rm sm}(R)/dR$ is the surface of a sphere with radius $r-R$, normalised by
the integral of the surface of spheres with radii from 0 to $r$: $dP_{\rm sm}(R)/dR=3 (r-R)^2/r^3$. 
If we define this as $W(r,R)$, then 
\beq
\frac{dP_{sm}}{dR}=
\int_R^\infty dr W(r,R)\frac{dP}{dr},
\eeq
is the bubble size distribution obtained by the
spherical average method for the real distribution $dP/dR$.  
The lower limit of the integral is $R$ because only spheres
which are larger than $R$ can contribute to the spherical average
bubble distribution at $R$ (because $x_{\rm th} \rightarrow 1$); 
the largest ionized sphere that can be
drawn around any given point is always smaller than or equal in radius
to the actual ionized sphere in which it lies.

\begin{figure}
\begin{center}
  \includegraphics[width=8cm]{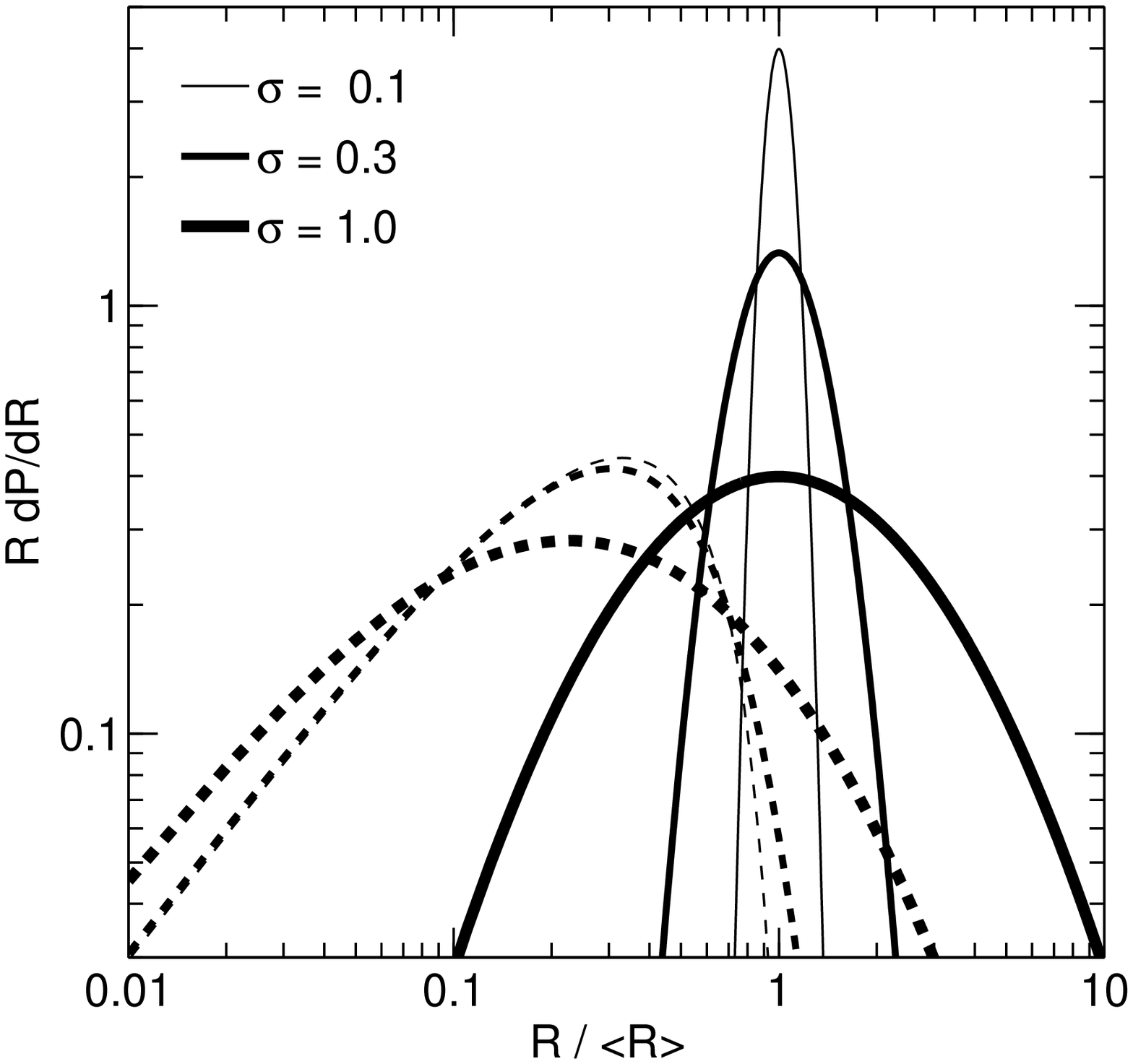}
  \caption{Spherical 
  average method applied to three log-normal distribution of non-overlapping 
  spherical H~II regions. The right (solid) curves are the actual log-normal 
  distributions given by equation (\ref{log-normal}) for different values of $\sigma$,
  while the left (dashed) curves show the result which would be obtained on the
  same distribution using the spherical average method.  As
  $\sigma\rightarrow 0$ and $dP/dR$ approaches a delta function,
  $dP_{sm}/dR$ approaches the kernel function $W(R,\langle R\rangle)$
\label{spa_test}
}
\end{center}
\end{figure}

Shown in Fig. \ref{spa_test} are $r dP/dr$ and the corresponding
$R dP_{sm}/dR$ for three log-normal distributions of bubble sizes
\beq
\frac{dP}{d\ln r}=\frac{1}{\sqrt{2\pi\sigma^2}}\exp - \left[
\frac{(\ln(r)-\ln(\langle r\rangle)^2}{2\sigma^2}\right],
\label{log-normal}
\eeq
with different $\sigma$.  As can be seen from the figure, the
spherical average tends to change the true bubble distribution in two
ways.  First, it smooths the actual bubble distribution with the
kernel $W(r,R)$. 
Second, it lowers the value of the mean bubble radius,
$R_{\rm av}=\int RdP/dR$.  In our simple toy model, the mean bubble size
obtained by the spherical average method is always 1/4 of the actual
one.              
Our toy model is admittedly crude, most notably in the assumption of a
threshold $x_{\rm th}=1$ and spherical H~II regions. A lower value of
$x_{\rm th}$ would allow small pockets of neutral gas to be attributed to
large ionized regions.  In fact, for the case where $x=1$ and $x=0$ in
ionized and neutral regions, respectively, lower values of $x_{\rm
  th}$ lead to an overestimate of the volume which is ionized, leading
to a violation of the normalization condition,
\beq
\int_0^\infty dR \frac{dP}{dR} = x_v.
\eeq
In most cases this overestimate is not very large.  A lower
value of $x_{\rm th}$ would also yield larger H~II regions, but this
effect has been shown to be rather modest \citep{2007ApJ...654...12Z}.  The
assumption of spherical symmetry is a conservative one, however, in the sense
that it provides a lower limit to how much the spherical average
method underestimates the ``true'' H~II regions sizes. 
This is because the spherical average method is sensitive to the smallest
dimension of the region in which it lies: the radius is
larger than the smallest dimension, the part of the sphere lying in
that direction would lie outside the region, and the average ionized
fraction would no longer be above the threshold for the region to be
considered ionized.

\section{Additional size Measure}
\label{MF_size_measure}
\begin{figure}
\begin{center}
  \includegraphics[width=8cm]{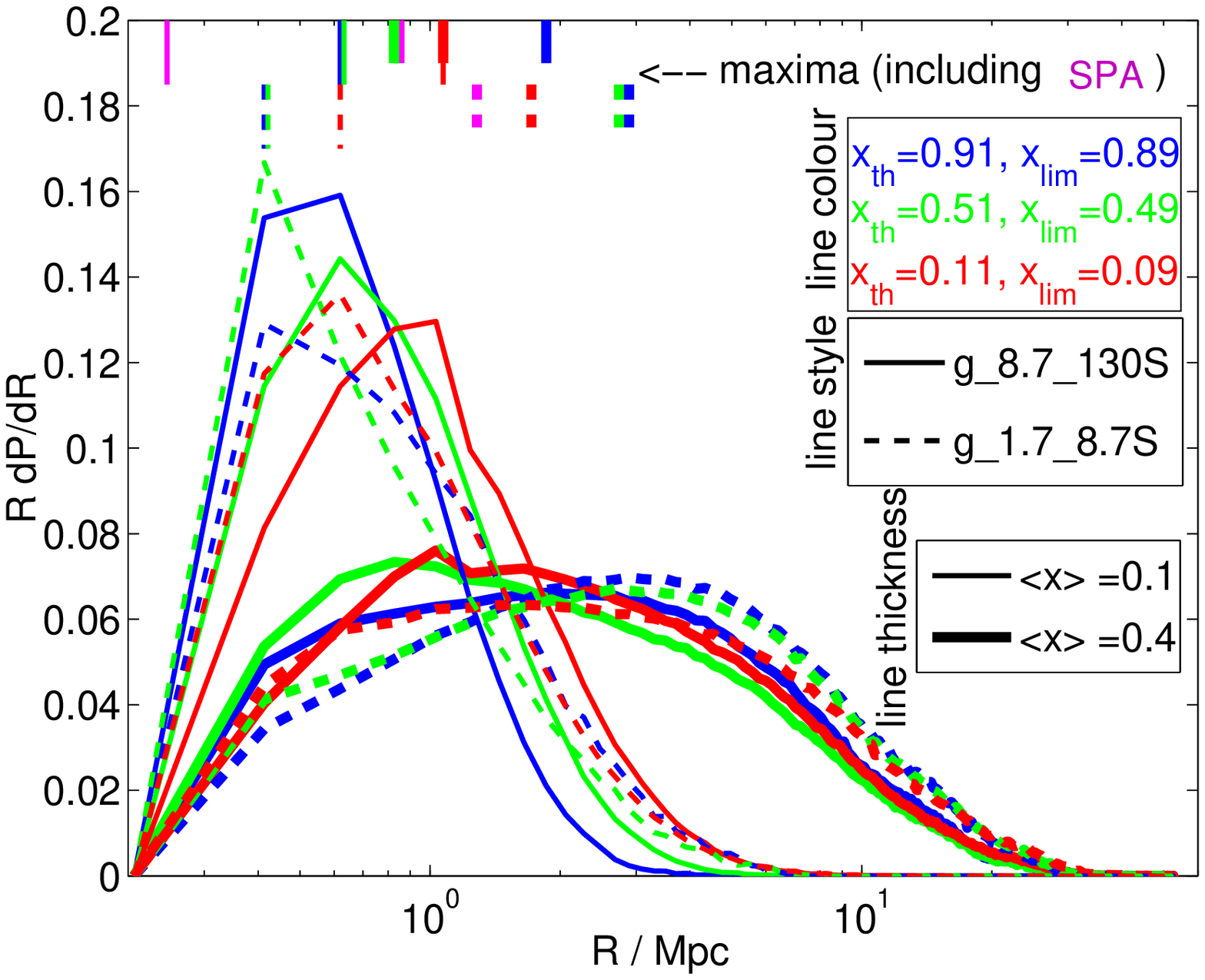}
  \caption{Size distribution curves for simulations $g8.7\_130S$ (solid lines) and $g1.7\_8.7S$ (dashed lines) using a method similar to the one described in \protect\cite{2007ApJ...669..663M} with three different combination of parameters $x_{\mathrm{th}}$ and $x_{\mathrm{lim}}$ (line colour) at two different global ionization fractions (line thickness). Also indicated are the maxima of the curves and the maxima of the SPA for the same simulations at the same global ionization fractions. Note the large effect of the choice of parameters $x_{\mathrm{th}}$ and $x_{\mathrm{lim}}$ on the position of the curve maximum. Note also the shift of the maxima of the SPA towards smaller scales. The maximum of SPA for simulation $g1.7\_8.7S$ at global ionization fraction $\left< x\right> \sim 0.1$ is at 0.1 Mpc and therefore not visible in this plot. 
\label{appendix0}
}
\end{center}
\end{figure}

\cite{2007ApJ...669..663M} used another technique to characterize sizes of 
ionized (and neutral) regions in their binary ionization fields: They chose a large number of random points; for each point they check if it is ionized (neutral); from each ionized (neutral) point they chose a random direction and measure the distance from this point to the nearest ionized-neural (neutral-ionized) transition boundary along that line of sight. 
In the following, we use a method similar to the one in \cite{2007ApJ...669..663M}.  However, instead of choosing a random direction, we, for simplicity, only choose between the 3 principle axis in one of the two possible orientations. The differences in size distribution obtained with this method only matters for very large and rare H~II regions. Small regions should be abundant enough that the orientation should not play a dominant role. Since we have to deal with continuos ionization fields, we have to introduce two parameters: Above which ionization fraction is a random point ionized? What is the limit for the transition boundary along the line of sight? In the following, we consider a point as ionized if its ionization fraction is above $x_{\rm{th}}$; we count each point along the line of sight ionized as long as its ionization fraction is above  $x_{\rm{lim}}$. In Fig.\ref{appendix0} we plot the size distribution curves obtained with this method for the two simulations we use in Section \ref{ana_method_intro}, $53Mpc\_g8.7\_130S$ (our fiducial simulation, solid lines) and $53Mpc\_g1.7\_8.7S$ (dashed lines) at two different global ionization fractions, $\left<x\right> \sim 0.1$ (thin lines) and $\left<x\right> \sim 0.4$ (thick lines). Qualitatively, we find the same result as with the other size measures: Initially the size distribution peaks earlier for $53Mpc\_g1.7\_8.7S$; at higher $\left<x\right>$, the typical size seems to be larger in this simulation than for the fiducial simulation. 

With respect to the peak position of the (uncorrected) SPA, especially at low global ionization fraction, the peak of this size distribution is shifted towards larger scales. This is consistent with the findings in appendix A and the fact that the size distribution of a single sphere, found with this method, would peak at the radius of the sphere; therefore, this method yields in theory better results than the SPA. However, the disadvantage is the dependence on two parameters if the field of investigation is not a binary field. We tested the effect of these parameters (compare different colours in Fig \ref{appendix0}) and find large variations for different parameter combinations.  

%-------------------------------------------------------------------------------
%-------------------------------------------------------------------------------
% EULER CHARACTERISTIC
\section{On the calculation of the Euler Characteristic}
\label{minkapp}
The estimation of $V_3$ of a three dimensional field sampled on a finite set of
 grid points is dependent on the chosen adjacency-pair, if the structure is of 
the same order as the cell sizes of the grid. An adjacency pair is for example 
(26,6) which means that cells above the threshold (foreground cells) have 26 
neighbours and cells below the threshold (background cells) have 6 neighbours.  
Because of this dependence on the choice of adjacency pair, we choose to 
oversample the data to avoid differences in the treatment of isolated/connected
 H~II and isolated/connected H~I regions. This introduces a stronger dependency
 on the choice of the threshold value. 

For calculating the Euler Characteristic $V_3$ we use part of a program developed
 by T. Buchert and J. Schmalzing \citep{1997ApJ...482L...1S}. The algorithm we 
use counts the vertices ($V$), edges ($E$), faces ($F$) and lattice cells ($C$) 
of the foreground cells and calculates $V_3$ according to equation (10) in \cite{1997ApJ...482L...1S}:
\beq
V_3= V - E + F - C
\eeq
In terms of adjacencies, this is equivalent to assigning 26 neighbor cells to any 
foreground cell and 6 neighbor cells to any background cell. This adjacency pair 
ensures the 3 dimensional Jordan curve theorem (which basically means that an 
edge-connection of foreground cells cannot at the same time be a connection 
for background cells). \cite{2002LNP...600..275O} showed that this adjacency 
pair is complementary which means that $V_3$ of the background is the same as 
that of the foreground. If the structure in the data-cube has contributions 
smaller or of the same size than is sampled by the grid-cells, structure of 
lower dimensions than 3 can arise. This can cause inconsistencies in the 
approximation of $V_3$ of the set sampled at the grid-points resulting in a 
violation of the complementarity. This is for example visible in the 
$(V_3,\delta_{th})$ plot in Fig. \ref{mink_dens}: The first peak is mainly 
due to disconnected under-dense regions which are below the density threshold 
value $\delta_{\rm th}$. If two of those regions are connected via an edge or 
a vertex, they count as two disconnected "cavities" since those cells are 
background cells and have only 6 neighbours. The second peak is somewhat smaller 
than the first one (although theoretically the peaks should be equal). This 
peak is mostly due to disconnected over-dense regions that are above the 
threshold value $\delta_{\rm th}$ and therefore have 26 neighbours each. 
Over-densities that are connected via an edge or a vertex count as one 
connected region, therefore their contribution to $V_3$ is smaller. 

\begin{figure}
\begin{center}
  \includegraphics[width=8cm]{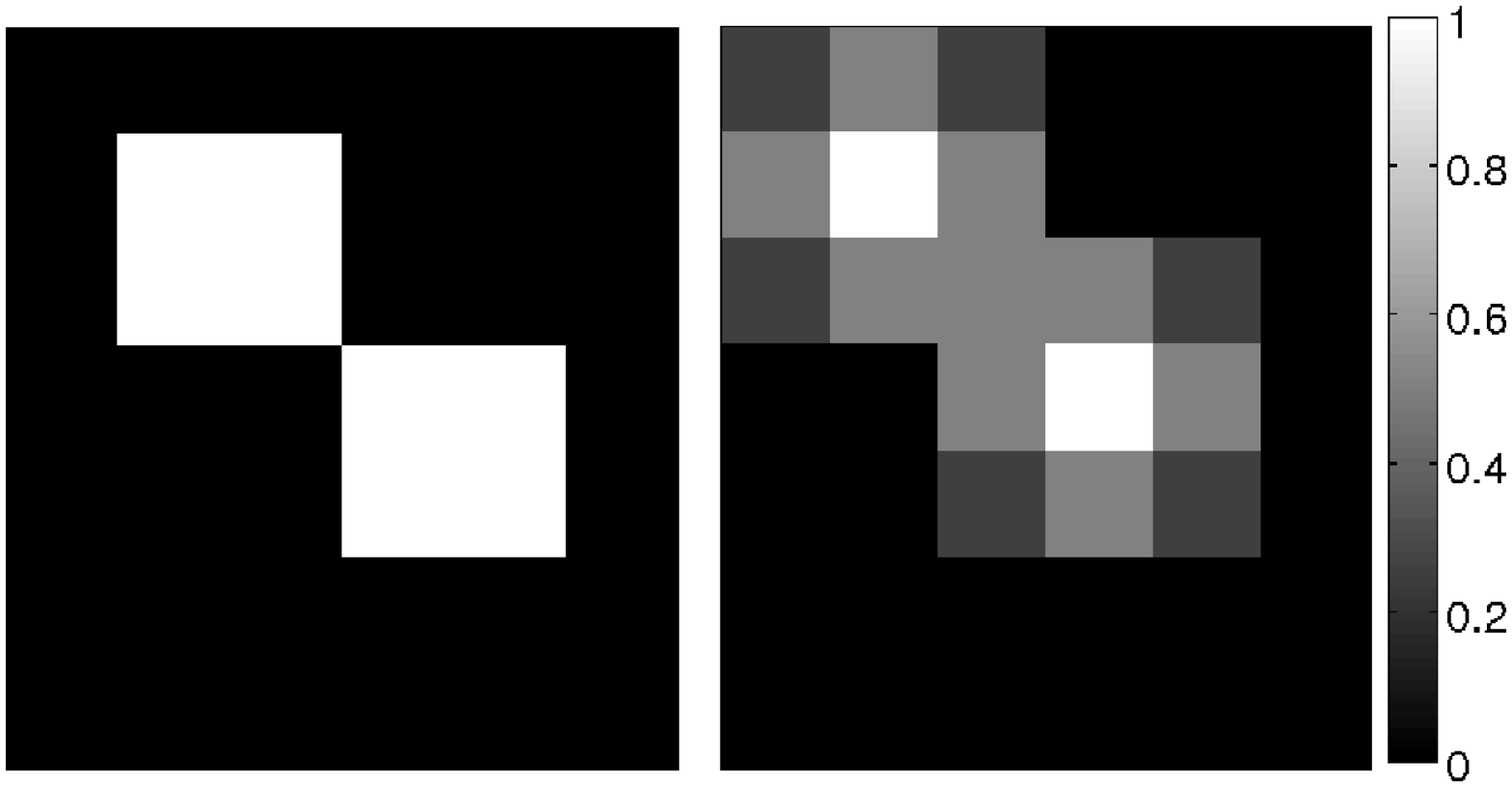}
  \caption{Two-dimensional example to demonstrate the effect of oversampling 
the data: original structure (left panel) compared to the result from 
oversampling (right panel). While the original structure counts as one
 connected region independent of threshold value ($x_{\rm th} \in (0,1)$),
 the structure in the right panel counts as two disconnected regions if 
$x_{\rm th}> 0.5$, and as one connected region otherwise. If the entries 
would be inversed (multiplied by -1), the structure in the left panel 
would count as two disconnected cavities while the structure in the right 
panel would have the same dependence on the threshold value as before: the 
cavities count as disconnected if $|x_{\rm th}|> 0.5$, and as one connected 
cavity otherwise. 
\label{appendix1}
}
\end{center}
\end{figure} 

\cite{2002LNP...600..275O} showed that the bias of the approximation depends 
on the choice of adjacencies. Inverting the data cube of ionization fraction,
 so that at every grid point, $x' = -x$, computing $V_3'$ at $x'_{\rm th} = -x_{\rm th}$
 and comparing $V_3'(x'_{\rm th})$ to $V_3(x_{\rm th})$,  is equivalent 
to changing the adjacencies from (26/6) to (6/26). 

For example, consider a  $3^3$ cube $C1$, with $C1(1,1,2)=C1(2,2,2)=1$ and
 0 everywhere else and a cube $C2=-C1$. Then,
 $V_3(C1,x_{\rm th}) \ne V_3(C2,-x_{\rm th})$: $V_3(C1,x_{\rm th}= l) = 14 - 23 + 12 -2  = 1$, 
where $l\in (0,1)$ (round brackets denoting open intervals) while 
$V_3(C2,x_{\rm th}=-l) = (\vartheta) - (\xi) + (\zeta) -(\varphi-2)  = 2$,
 where $\vartheta$, $\xi$, $\zeta $ and $\varphi $ are the numbers of vertices, 
edges, faces and lattice cells of the cube $C(1:3,1:3,1:3)=0$, the sum of which 
is 0 since periodicity is assumed. 

Therefore, we use $V_3(-x,-x_{\rm th})$ as a check for how good the approximation
 of $V_3$ is. To minimize the asymmetry due to the chosen adjacency, we oversample 
the data. For the example of the cubes $C1$ and $C2$ from above, this means 
that $V_3(C1,x_{\rm th}=l)=V_3(C2,x_{\rm th}=-l)= 1$ for $l \in (0,0.5)$ and
 $V_3(C1,x_{\rm th}=l)=V_3(C2,x_{\rm th}=-l)= 2$ for $l \in [0.5,1)$. Obviously
 this results in a higher dependency on the chosen threshold value, as this 
simple example demonstrates, see also Fig. \ref{appendix1} for a two-dimensional 
example.  

As a second example, we present here the Euler Characteristic of the fiducial
 simulation and simulation 53Mpc\_g1.7\_8.7S to demonstrate the effect of 
oversampling and the choice of threshold value. As explained above, calculating
 $V_3 (-x_{\rm th})$ of the inverted field (the data field multiplied by $-1$,
 in the following indicated by a ``-''), is equivalent to changing foreground 
to background cells and vice versa. This is the same as changing the adjacencies.
 Therefore, comparing $V_3 (+,x_{\rm th})$ to $V_3(-,-x_{\rm th})$ is a test 
for the dependence on adjacencies. 

\begin{figure}
\begin{center}
\includegraphics[width=8.1cm]{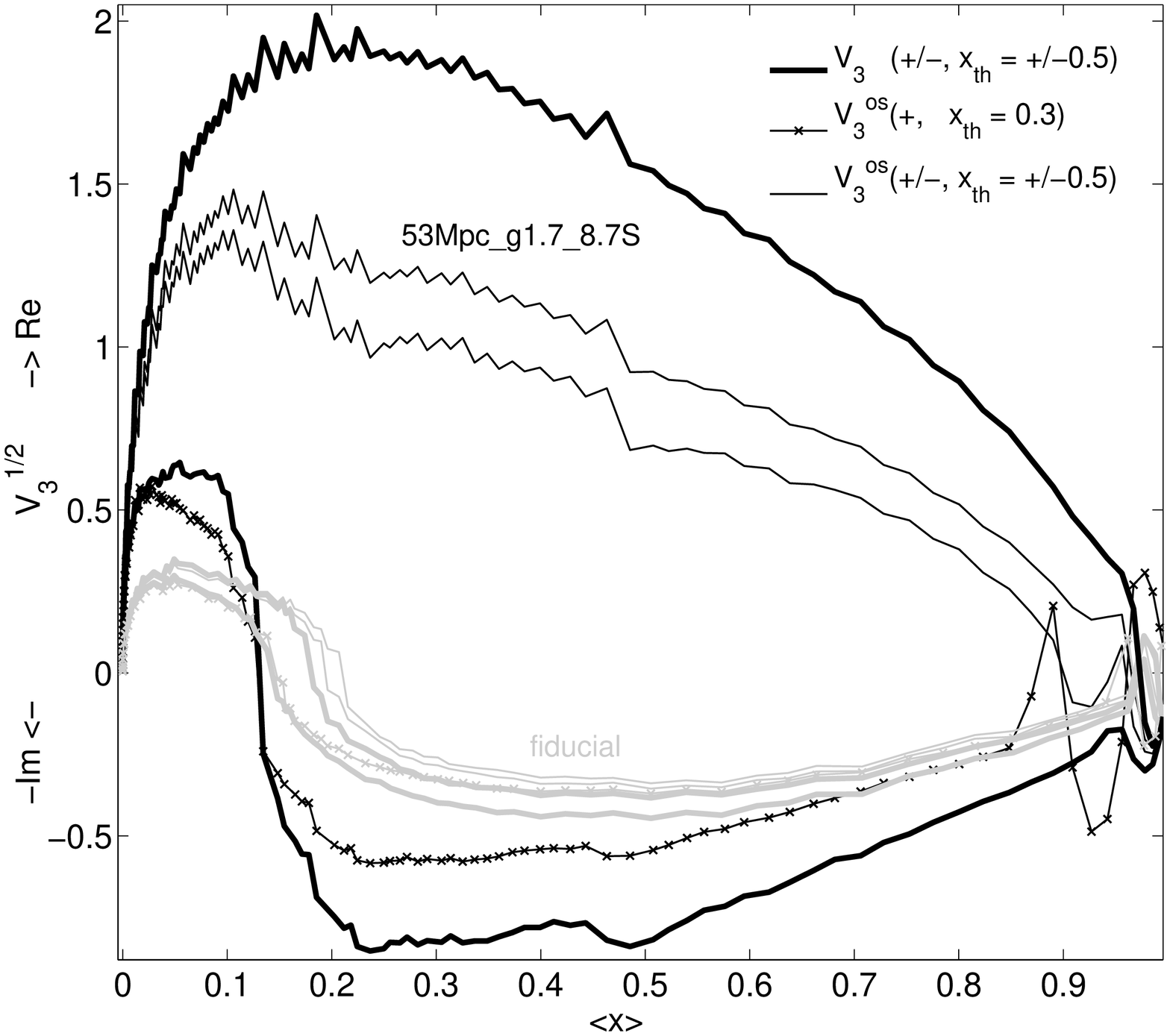}
\caption{Evolution of Euler Characteristic $V_3$ for the original field of
 the fiducial simulation (lower gray thick line)  and simulation 
53Mpc\_g1.7\_8.7S (lower black thick line) at threshold value  $x_{\rm th}=0.5$. 
The upper thick lines are the corresponding Euler Characteristic of the inversed 
fields at threshold value $x_{\rm th}=-0.5$. The thin lines correspond to the 
oversampled fields of the fiducial simulation (lower gray thin line) and
 simulation 53Mpc\_g1.7\_8.7S (lower black thin line) at threshold value  
$x_{\rm th}=0.5$. The upper thin lines are the corresonding Euler Characteristic
 of the inversed oversampled fields at threshold value $x_{\rm th}=-0.5$. The
 gray (fiducial simulation) and black (53Mpc\_g1.7\_8.7S) lines with crosses 
indicate the Euler Characteristic of the oversampled fields at threshold value 
$x_{\rm th}=0.3$. Note the good agreement for $V_3^{\rm os}(x_{\rm th}=0.5)$ 
and $V_3^{\rm os}(x_{\rm th}=-0.5)$ and the  low dependence on threshold value
 for the fiducial simulation. 
\label{xivx_b}
}
\end{center}
\end{figure}

Fig.\ref{xivx_b} demonstrates the dependence on the chosen adjacency pair 
when analyzing the data-fields without smoothing or oversampling (compare 
$V_3(+,x_{\rm th}=0.5)$ to $V_3(-,x_{\rm th}=-0.5)$ for both simulations) and
 the lower dependence on the chosen adjacencies after oversampling the data 
(compare $V_3^{\rm os}(+,x_{\rm th}=0.5)$ to $V_3^{\rm os}(-,x_{\rm th}=-0.5)$). 
It also demonstrates the relatively low dependence on threshold value $x_{\rm th}$
 for the fiducial simulation and the high dependence on  $x_{\rm th}$ for 
53Mpc\_g1.7\_8.7S (compare $V_3^{\rm os}(+,x_{\rm th}=0.5)$ to $V_3^{\rm os}(+,x_{\rm th}=0.3)$).

Others \citep[e.g.][]{2006MNRAS.370.1329G} chose to smooth the data with a 
gaussian kernel to remove lower-dimensional parts. This has the disadvantage 
that some of the small scale structures are suppressed. 
%-------------------------------------------------------------------------------
\end{appendix}
\label{lastpage}
\end{document}